\documentclass[twocolumn,aps,prl,reprint,superscriptaddress,longbibliography]{revtex4-2}
\usepackage{amssymb}
\usepackage{amsmath}
\usepackage{graphicx}
\usepackage{bbm}
\usepackage[dvipsnames, x11names]{xcolor}
\usepackage[colorlinks]{hyperref}
\hypersetup{
 colorlinks=true,
 citecolor=blue,
 linkcolor=blue,
 urlcolor=cyan}

\newcommand{\bra}[1]{\langle #1 \rvert}
\newcommand{\ket}[1]{\lvert #1 \rangle}

\begin{document}

\title{Efficient Simulation of Nonreciprocal Many-body Physics via Quantum Feedback}

\author{K. Vodenkova}
\thanks{These authors contributed equally.}
\affiliation{Institute for Theoretical Physics, University of Innsbruck, 6020 Innsbruck, Austria \\ and Institute for Quantum Optics and Quantum Information of the Austrian Academy of Sciences, \\ 6020 Innsbruck, Austria}

\author{A. Pocklington}
\thanks{These authors contributed equally.}
\affiliation{Pritzker School of Molecular Engineering, University of Chicago, Chicago, IL 60637, USA}

\affiliation{Department of Physics, University of Chicago, Chicago, IL 60637, USA}

\author{F. Yang}
\email{fan.yang@zju.edu.cn}
\affiliation{School of Physics and Zhejiang Key Laboratory of Micro-nano Quantum Chips \\ and Quantum Control, Zhejiang University, Hangzhou 310027, China}

\author{A. Lingenfelter}
\email{andrew.lingenfelter@uibk.ac.at}
\affiliation{Institute for Theoretical Physics, University of Innsbruck, 6020 Innsbruck, Austria \\ and Institute for Quantum Optics and Quantum Information of the Austrian Academy of Sciences, \\ 6020 Innsbruck, Austria}

\author{A. A. Clerk}
\affiliation{Pritzker School of Molecular Engineering, University of Chicago, Chicago, IL 60637, USA}

\author{H. Pichler}
\affiliation{Institute for Theoretical Physics, University of Innsbruck, 6020 Innsbruck, Austria \\ and Institute for Quantum Optics and Quantum Information of the Austrian Academy of Sciences, \\ 6020 Innsbruck, Austria}

%\date{\today}

\begin{abstract}
We present a scheme to efficiently simulate nonreciprocal many-body spin models using a non-Markovian open system. Our scheme utilizes a single quantum emitter coupled to a waveguide mode that is fed back to the same emitter after a time delay. With this coherent delayed feedback, an effective nonreciprocal interaction can be engineered, wherein the emitter at earlier times affects itself at later times, creating a scalable spin chain. We show that all the steady-state quantities of such spin models can be accessed through measurement of the emitter and the output field. Furthermore, using classical feedback that resets the emitter, quench dynamics from arbitrary product states can be simulated. We demonstrate striking features of nonreciprocal models, such as the Liouvillian skin effect, anomalous relaxation dynamics, and quasi-long-range order of output photons. Finally, we show that the proposal is amenable to experimental realization and robust to imperfections. Our work establishes a feasible way to scale up a nonreciprocal many-body system, and provides a new route towards generation of exotic states of light.
\end{abstract}

\maketitle

%%%%%%%%%%%%%%%%%%%%%%%%%%%%%%%%%%%%%%%%%%%%%%%%%%%

\textit{Introduction---}Driven-dissipative many-body systems often give rise to intriguing emergent phenomena absent in a fully closed setting. In particular, nonreciprocal models, where excitations propagate asymmetrically in opposite directions, have been actively studied in recent years \cite{Ashida2020,Bergholtz2021}. These studies not only broaden our knowledge of nonequilibrium many-body physics, such as the Liouvillian skin effect \cite{Haga2021} and new types of quantum criticality \cite{Begg2024,daraban_Universal_2026}, but also facilitate key applications such as quantum sensing \cite{Yang2023}, amplification \cite{metelmann2015nonreciprocal,wanjura2020topological,wanjura2021correspondence} and entanglement distribution \cite{PichlerPRA,stannigel_driven-dissipative_2012,irfan2024,irfan_Autonomous_2026}. Despite substantial theoretical advances, experimental realization of a large-scale, nonreciprocal many-body system remains challenging. For example, nonreciprocal couplings can be engineered using synthetic dimensions \cite{Ghatak2020,Wang2021,Wang2021_2,liang2022dynamic,Busnaina2024} or linear optics \cite{Weidemann2020,Xiao2021,weidemann2022topological}, but these approaches are often limited to the single-particle regime. Chiral waveguide quantum electrodynamics (QED) provides an alternative approach to constructing nonreciprocal spin models \cite{sollner2015deterministic,Joshi2023,Kannan2023}, but the chirality requires complex waveguide designs, while the inhomogeneity of artificial quantum emitters \cite{PhysRevLett.131.033606} makes it difficult to scale up the system.

Here we are interested in the physics of one-dimensional (1D) spin models in the extreme nonreciprocal regime, wherein excitations propagate in only one direction.
We demonstrate that these models can be efficiently simulated with a surprisingly simple setup:  a single physical quantum emitter (QE) coupled to a long waveguide which is terminated at one end by a mirror, with the resulting reflections creating coherent time-delayed feedback.
While coherent time-delayed feedback has been studied \cite{grimsmo_time-delayed_2015,whalen_open_2017,barkemeyer_Heisenberg_2022,crowder_quantum_2020,pichler_photonic_2016,zhang_Embedding_2022,regidor_qwavemps_2026} for applications such as quantum state control and stabilization \cite{carmele_Single_2013,Kraft_Timedelayed_2016,nemet_Stabilizing_2019,crowder_Improving_2024} and entanglement generation \cite{hein_Entanglement_2015,pichler_universal_2017,zou_Optimal_2026,ferreira_Deterministic_2024}, our scheme employs it to implement an analog quantum simulator by temporally multiplexing the many-body system using the single quantum emitter.
We show how phenomena characteristic of non-reciprocal many-body systems, such as the Liouvillian skin effect, asymmetric steady-state density distributions, and algebraic decay of correlations, manifest in our single-emitter setup.
We further discuss the effects of experimental imperfections and show the robustness of our scheme to realistic imperfections.
Our proposal can be readily implemented in existing superconducting circuit experiments  \cite{ferreira_Deterministic_2024,hoi_Probing_2015,mirhosseini_Superconducting_2018,ferreira_Collapse_2021} and is thus a promising scalable method to efficiently simulate a wide variety of nonreciprocal many-body models.

\begin{figure}
	\centering
	\includegraphics[width=\linewidth]{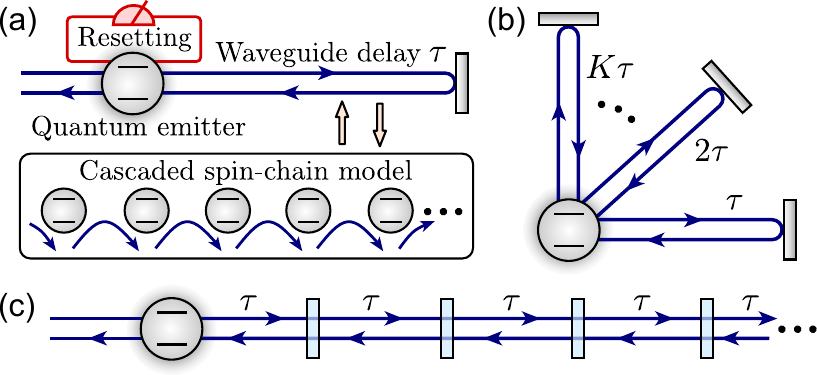}
	\caption{Illustration of the scheme and its generalized versions. (a) A single quantum emitter coupled to a 1D waveguide mode, subjected to coherent delayed feedback (blue lines) and incoherent measurement feedback (red lines). (b) A single quantum emitter coupled to multiple 1D waveguides, realizing up to $K$-local nonreciprocal couplings. (c) A single quantum emitter coupled to a 1D waveguide terminated by a Bragg mirror, mimicking a long-range interaction tail.
    } 
    \label{fig:model_intro} 
\end{figure}

\begin{table*}
	\caption{
	Correspondence between the effective many-body model and the original single-emitter system. In the final line, we use the notation $b_{{\rm out},n}(t) = b_{\rm out}[(n-1)\tau +t]$ for $t\in[0,\tau]$ \cite{SMarxiv}.} 
	\label{tab:table1}
	\begin{ruledtabular}
		\begin{tabular}{lcc}
			\multicolumn{1}{l}{\textrm{Physical quantities}}& 
			\multicolumn{1}{c}{\textrm{Effective $N$-body spin chain}}&
			\multicolumn{1}{c}{\textrm{Single QE system}}\\ 
			 \hline 
			Evolution time (for site $n$)   & $t \in [0, \tau] $ & $(n-1)\tau + t$ \\
            Separable initial state (with resetting) & $\rho_0 = \bigotimes_n\rho_0^{[n]}$ & $ \rho_\mathrm{QE}\left[(n-1)\tau\right]\overset{\rm reset}{\rightarrow}\rho_0^{[n]}$               \\
			Local observable  & $\langle{ O_n(t)}\rangle$   & $\langle  O[(n-1) \tau + t] \rangle$    \\
			Correlation function & $\langle  A_{n}(t)  B_{m}(t^\prime) \rangle$ & $ \langle  A[(n-1) \tau + t]  B[(m-1) \tau + t^\prime] \rangle$ \\
            Output field correlations & $\kappa\langle \sigma^+_n \sigma^-_m \rangle_t $ & $\sum_{j=0}^{n-1}\sum_{k=0}^{m-1} (-1)^{j+k}e^{\mathbbm{i}(k-j)\phi}\langle b_{{\rm out},n-j}^\dagger b_{{\rm out},m-k}\rangle_t$
		\end{tabular}
	\end{ruledtabular}
\end{table*}

\begin{figure}[b]
    \centering
    \includegraphics[width=\linewidth]{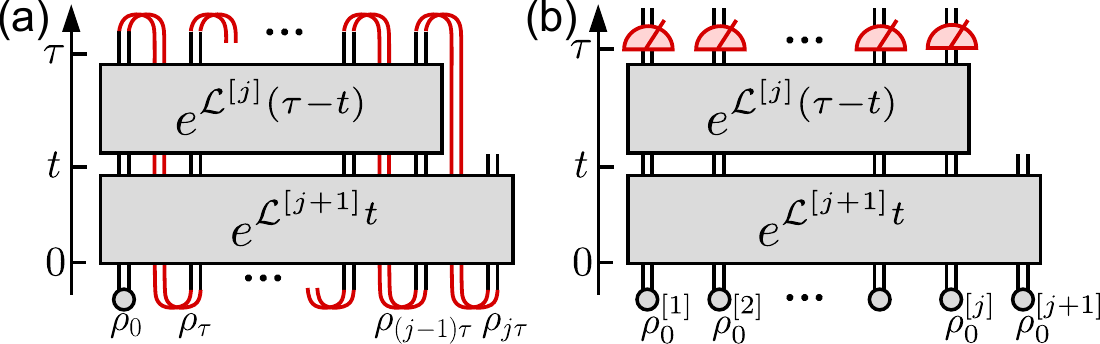}
    \caption{Quantum circuit illustration of the dynamics for the effective cascaded spin chain. (a) The full evolution corresponds to a shifted periodic boundary condition in which the initial state of each cascaded lattice site at $t=0$ is correlated with the final state at $t=\tau$ of the previous site (red tensor network connections). (b) Stroboscopic resetting can realize an arbitrary product initial many-body state.
    }
    \label{fig:twisted-boundary}
\end{figure}

%%%%%%%%%%%%%%%%%%%%%%%%%%%%%%%%%%%%%%%%%%%%%%%%%%%

{\textit{Simulating a nonreciprocal spin chain---}}Our starting point is the paradigmatic example of an $N$-site spin chain with nonreciprocal nearest-neighbor interaction:
\begin{align}
    \partial_t\rho &= -\mathbbm{i}[\sum_i(H_{{\rm s},i}+H_{i,i+1}),\rho] + \sum_i\mathcal{D}[L_{i,i+1}]\rho,  
    \label{eq:1d-cascaded-qme}
    \\
     H_{i,i+1} &= \frac{\mathbbm{i} \kappa}{2} \left( \sigma_i^+  \sigma_{i + 1}^- - \mathrm{h.c.}\right),\, 
     L_{i,i+1} = \sqrt{\kappa} \left( \sigma_i^- +  \sigma_{i+1}^- \right).
     \nonumber
\end{align}
Here $H_{{\rm s},i}$ is the local Hamiltonian on each lattice site $i$ (taken to be identical in the simplest case) and $\mathcal{D}[L]\rho = L\rho L^\dagger - \{L^\dagger L,\rho\}/2$ is the Lindblad dissipator.
The interplay between the Hamiltonian hopping and the dissipation causes excitation to hop from site $i$ to site $i+1$ but not in reverse, as captured by the effective non-Hermitian Hamiltonian $H_{\rm NH} = H - \frac{\mathbbm{i}}{2}\sum_i L_{i,i+1}^\dagger L_{i,i+1}$ that only contains the right-hopping terms $\propto\sigma_{i+1}^+\sigma_{i}^-$.

Our quantum simulator is a single quantum emitter (QE) subject to the local Hamiltonian $H_{\rm s}$ appearing in Eq.~\eqref{eq:1d-cascaded-qme} and coupled to a 1D waveguide with linear dispersion. 
We suppose a fraction of the emitted field is coherently fed back to the QE with a time delay $\tau$.
The prototypical setup is illustrated in Fig.~\ref{fig:model_intro}(a), where the coherent delayed feedback is enabled by field propagation and a perfectly reflecting mirror. Through this delayed feedback, the QE at time $t$ interacts effectively with itself at the later time $t+\tau$. Tracing over the field degrees of freedom, this setup can be exactly mapped to a cascaded spin chain comprising replicas of the QE at successive time $t,\,t+\tau,\,t+2\tau,\dots$ (see End Matter and Refs.~\cite{grimsmo_time-delayed_2015,Vodenkova2024}).
Thus, by evolving the single QE for a total time $N\tau$, we generate the effective dynamics of the $N$-site cascaded spin chain, evolved for a total time $\tau$.

Using this approach, each spin is by definition identical, thus, circumventing the challenge to scale up inhomogeneous artificial qubits to directly realize the many-body model Eq.~\eqref{eq:1d-cascaded-qme}.
Furthermore, the interaction is guaranteed to be nonreciprocal by causality, eliminating the need to engineer chirality in the waveguide. 
These advantages allow one to directly probe many-body nonreciprocal phenomena, which are beyond the capability of existing experiments limited to a small system size \cite{PRXQuantum.6.020101}.

Our general scheme is not limited to just realizing nearest-neighbor nonreciprocal couplings.
By introducing multiple feedback loops with commensurate time delays [see Figs.~\ref{fig:model_intro}(b) and \ref{fig:model_intro}(c)], it is possible to engineer nonlocal nonreciprocal couplings. Hence, our approach can simulate a rich class of open spin chains whose dynamics are governed by the master equation
\begin{equation}
\partial_t \rho=-\mathbbm{i}\left(H_{\rm NH}\rho-\rho H_{\rm NH}^\dagger\right)+\sum_{i,j}J_{ij}\sigma_{i}\rho \sigma_{j}^\dagger+\sum_{i,\mu}\mathcal{D}[L_{i}^\mu]\rho, \label{eq:master}
\end{equation}
where $\sigma$ is a generic spin operator of the QE that couples to the waveguide field. The non-Hermitian Hamiltonian $H_\mathrm{NH}$ induced by the coherent delayed feedback takes the nonreciprocal form
\begin{align}
    H_{\rm NH}&=\sum_{i} \left(H_{\mathrm{s},i}-\frac{\mathbbm{i}}{2}J_{ii}\sigma^\dagger_{i}\sigma_{i}\right)-{\mathbbm{i}}\sum_{i<j}J_{ij}\sigma^\dagger_{j}\sigma_{i}, \label{eq:general}
\end{align}
and the lattice indices ($i,j=1,2,\cdots, N$) correspond to the replica of the QE at different times.
Here, $L_i^\mu$ are local Markovian dissipation channels which can be engineered in the QE setup via standard reservoir engineering techniques, and the nonreciprocal interaction matrix $J_{ij}=J_{ji}^*$ depends on the configuration of feedback.

As an example of a setup with longer-range connectivity, additional delay loops with commensurate delay times $K\tau$, illustrated in Fig.~\ref{fig:model_intro}(b), realize nonreciprocal interactions $J_{i,i+K}\neq0$ between $K$-local sites, and a single Bragg mirror, shown in Fig.~\ref{fig:model_intro}(c), leads to a long-range tail whose amplitude $|J_{ij}|$ can mimic exponential $\alpha^{|i-j|}$ and power-law $|i-j|^\alpha$ decaying functions by optimizing the transmittance of each mirror \cite{SMarxiv}.
Finally, we note that the local Hamiltonians $H_{{\rm s},i}$ and dissipation channels $L_i^\mu$ in the cascaded chain are individually addressable by applying stroboscopic control of the QE Hamiltonian and its engineered reservoirs.

\begin{figure*}
	\centering
	\includegraphics[width=0.96\linewidth]{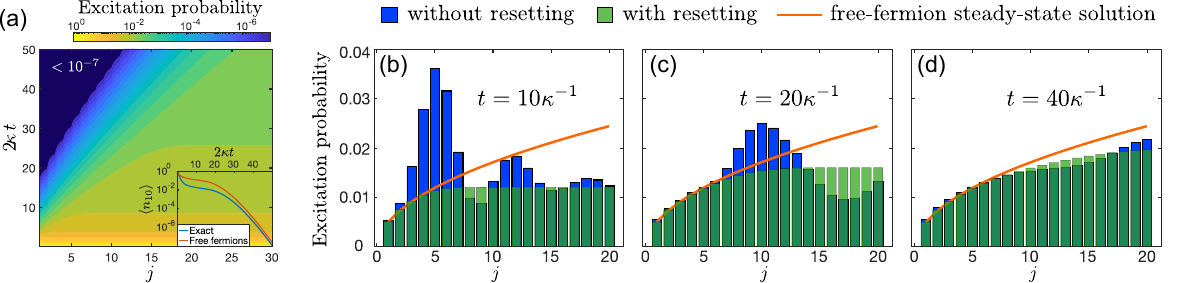}
	\caption{Evolution of local observables of the cascaded spin chain. (a) Excitation probability dynamics for a chain with only loss, with each QE initialized in the excited state.  The inset plot shows a comparison between the exact numerical calculation and free fermion approximation \cite{SMarxiv} for site $j=10$. (b)-(d) Excitation probability of spin chain sites $j$ under incoherent pumping at the indicated times $t=\tau$ for different delay times $\tau$. The green and blue bars show results with and without resetting, respectively. The solid lines show the steady-state solutions under the free-fermion approximation. For all plots, the numerical simulation Trotter step is $2\kappa dt=0.001$ and for (b)-(d) the incoherent pump strength is $\Gamma=0.01\kappa$.
    }
    \label{fig:1dcasc_ss}
\end{figure*}

%%%%%%%%%%%%%%%%%%%%%%%%%%%%%%%%%%%%%%%%%%%%%%%%%%%

{\textit{Simulation strategies---}}To use the QE setup as an analog simulator of the many-body physics governed by Eq.~\eqref{eq:master}, 
we must determine how to prepare a desired initial many-body state $\rho_0$ of the spin chain into our single QE simulator and establish how to measure  observables and correlation functions.
The $j=1$ site of the spin chain is clearly initialized in the initial state of the QE; however, without further intervention, the initial state of subsequent sites $j>1$ depend on the state of the QE at time $j\tau$, as illustrated in Fig.~\ref{fig:twisted-boundary}(a).
This generally implies complicated multi-body correlations in the effective spin chain initial state. Nevertheless, as long as the feedback delay time $\tau$ is sufficiently large (assuming a unique steady state and non-zero dissipative gap), the correlations of the initial many-body state decay away and the effective spin chain relaxes to its steady state. 
As a result, all the steady-state observables become accessible, e.g., the two-site correlation $\langle A_{n}  B_{m}\rangle_\mathrm{ss}$ of the effective spin chain is identical to the two-time correlation $\langle A(n\tau)  B(m\tau)\rangle$ of the QE as $\tau\rightarrow \infty$.

Our scheme is not restricted to steady-state observables. 
It also allows one to simulate finite-time quench dynamics from a separable initial state $\rho_0^{[1]}\otimes\rho_0^{[2]}\otimes\cdots\rho_0^{[N]}$. 
This can be achieved via a fast reset procedure: at time $j\tau$, one rapidly resets the QE to $\rho_0^{[j+1]}$, e.g., via a projective measurement followed by a conditional state-preparation operation that prepares the QE in the target state, see Fig.~\ref{fig:twisted-boundary}(b). For a finite resetting time $t_r$, the maximum evolution time that can be accessed for the effective model decreases to $\tau-t_r$. To retain an almost perfect initialization, we need to ensure $t_r\ll\kappa^{-1}$ or turn off the QE-waveguide coupling during the resetting \cite{SMarxiv}.

Local observables and correlation functions of the spin chain at finite times $t < \tau$ are accessible in the single QE setup as outlined in the dictionary Table \ref{tab:table1}.
For local observables, it is convenient to directly probe the QE at the times indicated in Table \ref{tab:table1}. Measurement of correlation functions can be more complicated, as it involves evaluation of multi-time correlations of the QE. Fortunately, the QE-waveguide coupling allows us to extract certain correlation functions by probing the output photonic field determined by the input-output relation, e.g., for the setup shown in Fig.~\ref{fig:model_intro}(a), $b_\mathrm{out}(t) = \sqrt{\kappa}[\sigma^{-}(t) + e^{-\mathbbm{i}\phi}\sigma^{-}(t-\tau)] + e^{-\mathbbm{i}\phi}b_{\rm in}(t)$, where $\phi = \pi - \omega_0 \tau$ is the phase acquired by a photon during a round-trip in the delay loop and $b_{\rm in}(t)$ is the input field. By measuring a set of coherence \cite{astafiev2010resonance} or intensity correlations \cite{arcari2014near} between output photons, it is possible to reconstruct spin-spin correlations of the effective model, as given, for example, in the final row of Table~\ref{tab:table1}.

%%%%%%%%%%%%%%%%%%%%%%%%%%%%%%%%%%%%%%%%%%%%%%%%%%%

{\textit{1D cascaded network example---}}
The simplest example of the nearest-neighbor cascaded chain [Eq.~\eqref{eq:1d-cascaded-qme}] already exhibits intriguing observable phenomena.
Such a system was previously studied in Ref.~\cite{Begg2024} (and a free-particle version in \cite{McDonald2022}), where it has been observed that the critical nature of the model gives rise to anomalous power-law relaxation dynamics of the local spin operators, which is expected to be different from the reciprocal case \cite{Begg2024}. 

We numerically simulate this model using the single QE setup in Fig.~\ref{fig:model_intro}(a) with $H_{\rm s}=0$ via the tensor network technique of Ref.~\cite{Vodenkova2024}.
To observe signatures of the nonreciprocity, we first consider the quench dynamics of a fully excited initial state, which can be realized by resetting the QE to the excited state at each time $n\tau$. 
As shown in Fig.~\ref{fig:1dcasc_ss}(a), the global relaxation time is on the order of $N \kappa^{-1}$, much longer than that set by the sole frequency scale $\kappa$.
This is a direct manifestation of the Liouvillian skin effect (LSE) \cite{Yao2018,Okuma2020,zhang2022review,lee_Anomalously_2023}, in which nonreciprocity causes all eigenmodes of a system with open boundary conditions to be exponentially localized to a single boundary, giving rise to an anomalously long relaxation time despite the existence of a finite Liouvillian gap \cite{Haga2021}.
During the relaxation, there are two competing effects we can observe: firstly, after a short transient dynamics, the local excitation probability follows an algebraic decay in time, which is related to the fact that the system has a gapless pseudospectrum \cite{trefethen2005}; secondly, after a time that scales with system size, a wavefront passes through the spin chain after which the decay becomes exponential again.

The nonreciprocity can also manifest itself in steady-state properties. 
To achieve a nontrivial steady state, we introduce a weak incoherent pump to the QE by adding local Lindblad terms
\begin{align}
     L_{i}^\mathrm{pump} &= \sqrt{\Gamma} \sigma_i^+
\end{align}
to the effective spin chain. 
A weak pump $\Gamma\ll\kappa$ slightly pushes the system away from criticality, but on length scales shorter than $\xi_L = \kappa/\Gamma$, the model still exhibits critical behaviors altered by the nonreciprocity. 
For example, we find that the steady-state spin density $\langle \sigma_i^z \rangle_{\mathrm{ss}}$ exhibits a highly asymmetric distribution.
We estimate this distribution by making a free-fermion approximation: we make a Jordan-Wigner transformation of Eq.~\eqref{eq:1d-cascaded-qme} and drop the fermion strings appearing in the dissipators to obtain a noninteracting fermion model whose steady state distribution is given by \cite{SMarxiv}
\begin{align}
    \langle \sigma_i^z \rangle_{\mathrm{ss}} & \sim -1 + \frac{4 \xi_L^{-1}}{\sqrt{\pi} } \sqrt{i} + \mathcal{O}\left( \xi_L^{-2} \right).
\end{align}
Note that the free-fermion approximation is uncontrolled, so one should not expect precise quantitative agreement between it and the spin model.
To probe such a steady-state density profile, resetting of the emitter is not required, as long as the feedback delay time is sufficiently large. 
This is demonstrated in Fig.~\ref{fig:1dcasc_ss}(b)-(d), where we show the excitation probability $\langle\sigma_i^+\sigma_i^-\rangle$ at the final evolution time, $t=\tau$ for varying delay times $\tau$. 
Despite the appearance of oscillations in intermediate time scales, the density profile in the non-resetting regime eventually converges to the same pattern as the case with resetting for sufficiently long $\tau$. 
They both reveal the algebraic growth predicted by the free-fermion approximation, with deviations appearing for sites at the far end of the chain.
these deviations do not represent a failing of our simulation scheme as the uncontrolled free-fermion approximation is a priori not expected to quantitatively agree.

%%%%%%%%%%%%%%%%%%%%%%%%%%%%%%%%%%%%%%%%%%%%%%%%%%%

\begin{figure}[b]
	\centering
	\includegraphics[width=\linewidth]{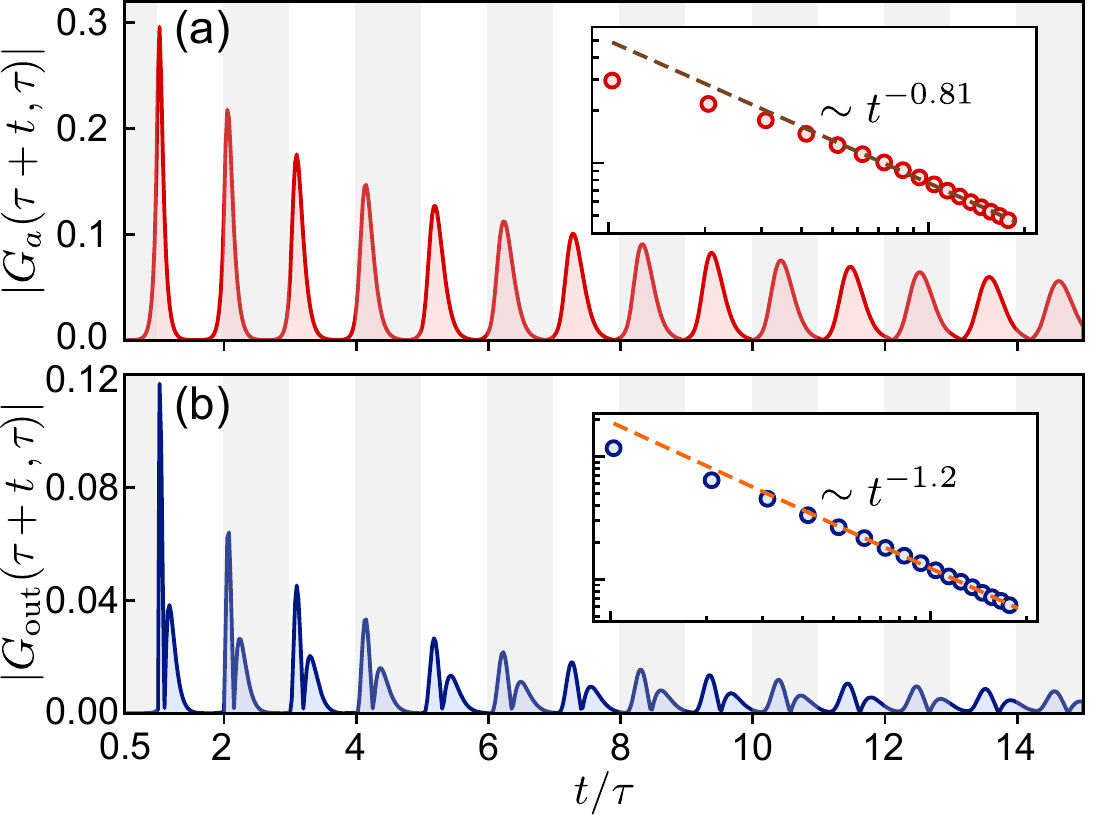}
	\caption{Two-time correlation functions of the quantum emitter under the delayed feedback and a weak incoherent pump (without resetting). (a) Magnitude of atomic correlations $G_a(\tau+t,\tau) = \kappa\langle \sigma^+(\tau+t) \sigma^-(\tau)\rangle/\Gamma$. (b) Magnitude of output field correlations $G_\mathrm{out}(\tau+t,\tau) = \langle b_\mathrm{out}^\dagger(\tau+t) b_\mathrm{out}(\tau)\rangle/\Gamma$. The insets show the power-law fitting to peak values of the correlations. The parameters are the same as in Fig.~\ref{fig:1dcasc_ss}(d). The shading bars indicate lattice sites in the spin chain. E.g., for $t/\tau\in[4,5]$ the many-body correlation is $\mathcal{G}_{6,2}(t',0)=\langle\sigma_6^+(t')\sigma^-_2(0)\rangle$ for $t'\in[0,\tau]$.
    }
    \label{fig:correlation}
\end{figure}

{\textit{Probing the quasi-long-range order---}}The near-critical feature of the system under a weak incoherent pump can give rise to nontrivial quantum correlations, such as the algebraic decay of certain correlation functions. 
To probe such a quasi-long-range order, we first examine the atomic two-time correlation function $G_a(\tau+t,\tau) = \kappa\langle \sigma^+(\tau+t)\sigma^-(\tau)\rangle/\Gamma$, where $\tau$ is chosen to be sufficiently large to let the effective spin chain reach the steady state. 
This correlation function maps to the many-body two-time correlation $\mathcal{G}_{n,2}(t',0)=\langle \sigma_n^+(t')\sigma_2^-(0)\rangle$ between site 2 at $t_0=0$ and site $n = 2 + \lfloor t/\tau \rfloor$ at the latter time $t' = t - \lfloor t/\tau \rfloor$
(Equivalently, this can be defined as as $\mathcal{G}_{n,1}(t',\tau) = \langle\sigma_n^+(t')\sigma_1(\tau)\rangle$, with $n=1+\lfloor t/\tau\rfloor$.)
As shown in Fig.~\ref{fig:correlation}(a), the correlation $G_a(\tau+t,\tau)$ displays the collapse and revival (without resetting), whose peaks are equally spaced at a time interval $\sim \tau+(1/\kappa)$ and indeed exhibit power-law scaling with the duration $t$.
The spacing of the peaks implies that the correlations between sites distance $d$ apart do not occur at equal times in the many-body system, but at time differences $\sim d/\kappa$.

In addition to atomic correlations, one can also directly probe the two-time output field correlation $G_\mathrm{out}(\tau+t,\tau) = \langle b_\mathrm{out}^\dagger(\tau+t) b_\mathrm{out}(\tau)\rangle/\Gamma$.
This corresponds to the spin-chain output-field correlation function $\langle b_{{\rm out},n}^\dagger(t')b_{{\rm out},2}(0)\rangle$ for $n = \lfloor t/\tau\rfloor + 2$ and $t' = t-\lfloor t/\tau\rfloor$ (see Table~\ref{tab:table1}).
Interestingly, the quasi-long-range order is preserved in the output field, but the destructive interference between the emissions at distinct times modifies the power-law exponent [see Fig.~\ref{fig:correlation}(b)].
The double-peak of $G_{\rm out}(\tau+t,\tau)$ at each interval is a correlation ($G_{\rm out}>0$) followed immediately by an anti-correlation ($G_{\rm out}<0$), but in the figure we plot the absolute value.
We note that such a correlated photonic many-body state is highly nontrivial, as it differs from either the short-range correlated field $G_\mathrm{out}(\tau+t,\tau)\propto e^{-(\kappa+\Gamma/2)t}$ emitted by a bare, incoherently pumped emitter, or the true-long-range order $G_\mathrm{out}(\tau+t,\tau)\rightarrow \mathit{const}$ established by a fully coherent laser driving. 
Furthermore, we find that while the free-fermion approximation reveals similar power-law behaviors \cite{SMarxiv}, it yields different power-law exponents, implying that the many-body effects can play a significant role in the case of spin dynamics, as is also observed in Refs.~\cite{Begg2024, Pocklington2025}.

%%%%%%%%%%%%%%%%%%%%%%%%%%%%%%%%%%%%%%%%%%%%%%%%%%%

{\textit{Experimental considerations---}}The proposal is ideally suited for a number of experimental platforms. 
Here we consider a superconducting qubit coupled to a microwave metamaterial waveguide with a very high index of refraction to create long delay times \cite{mirhosseini_Superconducting_2018,ferreira_Collapse_2021}. 
Such an experiment must contend with parasitic dissipation and nonlinear dispersion.

\begin{figure}[t]
	\centering
	\includegraphics[width=0.99\linewidth]{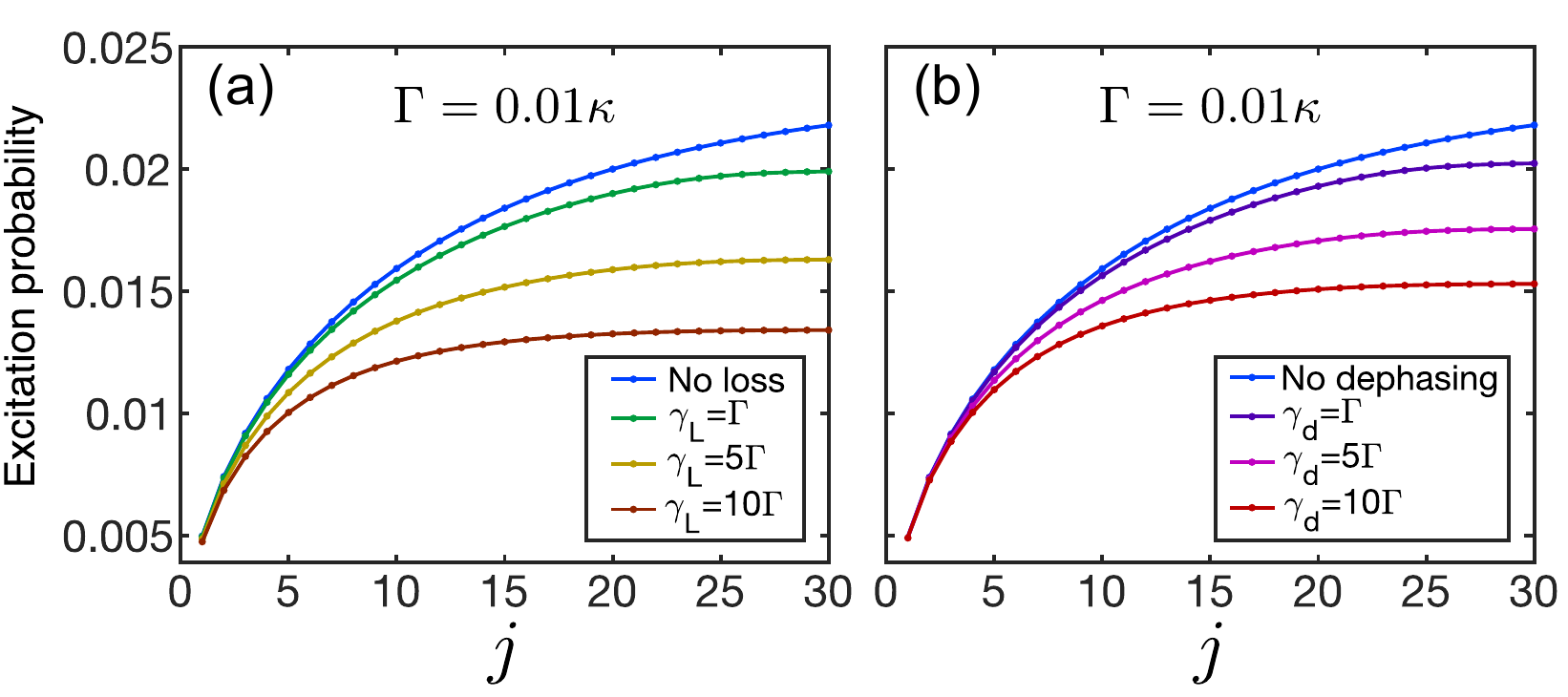}
	\caption{Steady state properties of the 1D cascaded chain [Eq.~\eqref{eq:1d-cascaded-qme}] with the effects of the excitation loss (a) or dephasing (b) with the incoherent pump strength $\Gamma=0.01\kappa$.
    The simulation Trotter step is $2\kappa dt=0.001$, and the delay time is $\tau = 50\kappa^{-1}$, which is sufficient to approximate steady state.
    }
    \label{fig:loss_effects} 
\end{figure}

The primary sources of parasitic dissipation are intrinsic qubit excitation loss, photon loss in the waveguide, and qubit dephasing.
Waveguide loss effectively induces additional loss on each qubit \cite{irfan2024}, hence all parasitic dissipation can be included by introducing a local loss channel $L_{i}^\mathrm{loss} = \sqrt{\gamma_{\rm L}}  \sigma_i^-$ and a local dephasing channel $L_{i}^\mathrm{deph} = \sqrt{\gamma_{\rm d}}  \sigma^+_{i} \sigma^-_{i}$ 
into Eq.~\eqref{eq:1d-cascaded-qme}. 
We show in Fig.~\ref{fig:loss_effects} that the algebraic scaling of the spin density persists over a length scale set by the parasitic dissipation rates \cite{SMarxiv}. 
Thus, the near-critical behavior is robust to experimental imperfections.

Reaching the steady state of a length-$N$ chain, requires a delay time $\kappa \tau \gtrsim  N$. The maximum simulable chain length $N$ is primarily constrained by the combined parasitic loss rate $\gamma_{\rm L}$ (assuming the dephasing rate to be small in comparison), $N\tau\gamma_{\rm L}\ll1$. 
When combined with the long delay time requirement, this yields a maximum simulable chain length $N_{\rm max} \sim \sqrt{{\kappa}/{\gamma_{\rm L}}}$.
In a superconducting circuit, structured metamaterial waveguides which have a group index approaching $n_g \sim 1000$ near a band edge have been experimentally realized \cite{mirhosseini_Superconducting_2018}.
Coupling a transmon qubit operating at 4.8 GHz to one end of an approximately ${12~\text{cm}}$ section of the metamaterial waveguide designed in Ref.~\cite{mirhosseini_Superconducting_2018} with a reflective termination at the other end, we find that the steady-state properties of a length $N_{\rm max} \approx 10$ chain can be probed.
The simulable chain length is limited by both waveguide loss and dispersion \cite{SMarxiv}; we believe this length can be increased by optimizing the metamaterial design.

%%%%%%%%%%%%%%%%%%%%%%%%%%%%%%%%%%%%%%%%%%%%%%%%%%%

{\textit{Conclusion and outlook---}}We have proposed a novel analog quantum simulation method for implementing driven-dissipative many-body systems with fully nonreciprocal interactions by temporally multiplexing the many-body dynamics using a single delayed-feedback-controlled quantum emitter.
We demonstrate that our scheme simulates the observable phenomena characteristic of a nonreciprocal model, such as the anomalously slow relaxation dynamics induced by the Liouvillian skin effect and emergent quasi-long-range order of the output photonic field.
The scheme is robust to imperfections such as nonlinear dispersion effects, excitation losses, and dephasing, which makes it promising for near-term experimental realizations.

While our scheme offers a way to study the fully nonreciprocal regime, it remains an interesting open question whether a tunable nonreciprocity can be engineered, e.g., through retrodiction measurements \cite{Gammelmark2013past,bao2020spin}. Our model can be further generalized to more complicated networks by including multiple feedback loops or quantum emitters to explore the physics of long-range nonreciprocal interactions \cite{agusti_Autonomous_2023}.

{\textit{Acknowledgments---}}We thank Gideon Lee, Bo Yan, Yuxiang Zhang, Yaohua Li, A. Rex, and S. White for valuable discussions. 
We acknowledge support from the Army Research Office via grant number W911NF-25-1-0286, and the Simons Foundation through a Simons Investigator Award (Grant No. 669487). This work was also supported by the European
Union’s Horizon Europe research and innovation program
under Grant No. 101113690 (PASQuanS2.1), the ERC
Starting grant QARA (Grant No. 101041435), the Austrian
Science Fund (FWF) (Grant No. DOI 10.55776/COE1).

\bibliography{references}

@misc{SMarxiv,
	note = {See Supplemental Material for details. 
    }
}

@article{metelmann2015nonreciprocal,
  title = {Nonreciprocal Photon Transmission and Amplification via Reservoir Engineering},
  author = {Metelmann, A. and Clerk, A. A.},
  journal = {Phys. Rev. X},
  volume = {5},
  issue = {2},
  pages = {021025},
  numpages = {16},
  year = {2015},
  month = {Jun},
  publisher = {American Physical Society},
  doi = {10.1103/PhysRevX.5.021025}
}

@article{wanjura2020topological,
  title={Topological framework for directional amplification in driven-dissipative cavity arrays},
  author={Wanjura, Clara C and Brunelli, Matteo and Nunnenkamp, Andreas},
  journal={Nat. Commun.},
  volume={11},
  number={1},
  pages={3149},
  year={2020},
  publisher={Nature Publishing Group UK London},
  doi={10.1038/s41467-020-16863-9}
}

@article{wanjura2021correspondence,
  title = {Correspondence between Non-Hermitian Topology and Directional Amplification in the Presence of Disorder},
  author = {Wanjura, Clara C. and Brunelli, Matteo and Nunnenkamp, Andreas},
  journal = {Phys. Rev. Lett.},
  volume = {127},
  issue = {21},
  pages = {213601},
  numpages = {6},
  year = {2021},
  month = {Nov},
  publisher = {American Physical Society},
  doi = {10.1103/PhysRevLett.127.213601}
}

@article{stannigel_driven-dissipative_2012,
	title = {Driven-dissipative preparation of entangled states in cascaded quantum-optical networks},
	volume = {14},
	issn = {1367-2630},
	url = {https://iopscience.iop.org/article/10.1088/1367-2630/14/6/063014},
	doi = {10.1088/1367-2630/14/6/063014},
	number = {6},
	urldate = {2023-01-04},
	journal = {New J. Phys.},
	author = {Stannigel, K and Rabl, P and Zoller, P},
	month = jun,
	year = {2012},
	pages = {063014},
}

@article{zhang2022review,
  title={A review on non-Hermitian skin effect},
  author={Zhang, Xiujuan and Zhang, Tian and Lu, Ming-Hui and Chen, Yan-Feng},
  journal={Adv. Phys.},
  volume={7},
  number={1},
  pages={2109431},
  year={2022},
  publisher={Taylor \& Francis},
  doi={10.1080/23746149.2022.2109431}
}

@article{PRXQuantum.6.020101,
  title = {Chiral Quantum Optics: Recent Developments and Future Directions},
  author = {Su\'arez-Forero, D.G. and Jalali Mehrabad, M. and Vega, C. and Gonz\'alez-Tudela, A. and Hafezi, M.},
  journal = {PRX Quantum},
  volume = {6},
  issue = {2},
  pages = {020101},
  numpages = {18},
  year = {2025},
  month = {Apr},
  publisher = {American Physical Society},
  doi = {10.1103/PRXQuantum.6.020101},
  url = {https://link.aps.org/doi/10.1103/PRXQuantum.6.020101}
}

@article{Vodenkova2024,
  title = {Continuous Coherent Quantum Feedback with Time Delays: Tensor Network Solution},
  author = {Vodenkova, Kseniia and Pichler, Hannes},
  journal = {Phys. Rev. X},
  volume = {14},
  issue = {3},
  pages = {031043},
  numpages = {32},
  year = {2024},
  month = {Sep},
  publisher = {American Physical Society},
  doi = {10.1103/PhysRevX.14.031043},
  url = {https://link.aps.org/doi/10.1103/PhysRevX.14.031043}
}

@article{Kraft_Timedelayed_2016,
  title = {Time-delayed quantum coherent Pyragas feedback control of photon squeezing in a degenerate parametric oscillator},
  author = {Kraft, Manuel and Hein, Sven M. and Lehnert, Judith and Sch\"oll, Eckehard and Hughes, Stephen and Knorr, Andreas},
  journal = {Phys. Rev. A},
  volume = {94},
  issue = {2},
  pages = {023806},
  numpages = {11},
  year = {2016},
  month = {Aug},
  publisher = {American Physical Society},
  doi = {10.1103/PhysRevA.94.023806},
  url = {https://link.aps.org/doi/10.1103/PhysRevA.94.023806}
}

@article{PichlerPRA,
  title = {Quantum optics of chiral spin networks},
  author = {Pichler, Hannes and Ramos, Tom\'as and Daley, Andrew J. and Zoller, Peter},
  journal = {Phys. Rev. A},
  volume = {91},
  issue = {4},
  pages = {042116},
  numpages = {19},
  year = {2015},
  month = {Apr},
  publisher = {American Physical Society},
  doi = {10.1103/PhysRevA.91.042116},
  url = {https://link.aps.org/doi/10.1103/PhysRevA.91.042116}
}

@article{pichler_universal_2017,
	title = {{Universal Photonic Quantum Computation Via Time-Delayed Feedback}},
	volume = {114},
	issn = {0027-8424, 1091-6490},
	url = {https://pnas.org/doi/full/10.1073/pnas.1711003114},
	doi = {10.1073/pnas.1711003114},
	number = {43},
	urldate = {2022-05-12},
	journal = {Proc. Natl. Acad. Sci. U.S.A.},
	author = {Pichler, Hannes and Choi, Soonwon and Zoller, Peter and Lukin, Mikhail D.},
	month = oct,
	year = {2017},
	pages = {11362--11367},
}

@article{carmele_Single_2013,
  title = {{Single Photon Delayed Feedback: A Way to Stabilize Intrinsic Quantum Cavity Electrodynamics}},
  author = {Carmele, Alexander and Kabuss, Julia and Schulze, Franz and Reitzenstein, Stephan and Knorr, Andreas},
  journal = {Phys. Rev. Lett.},
  volume = {110},
  issue = {1},
  pages = {013601},
  numpages = {5},
  year = {2013},
  month = {Jan},
  publisher = {American Physical Society},
  doi = {10.1103/PhysRevLett.110.013601},
  url = {https://link.aps.org/doi/10.1103/PhysRevLett.110.013601}
}

@article{nemet_Stabilizing_2019,
  title = {Stabilizing quantum coherence against pure dephasing in the presence of time-delayed coherent feedback at finite temperature},
  author = {N\'emet, Nikolett and Parkins, Scott and Knorr, Andreas and Carmele, Alexander},
  journal = {Phys. Rev. A},
  volume = {99},
  issue = {5},
  pages = {053809},
  numpages = {12},
  year = {2019},
  month = {May},
  publisher = {American Physical Society},
  doi = {10.1103/PhysRevA.99.053809},
  url = {https://link.aps.org/doi/10.1103/PhysRevA.99.053809}
}

@article{hein_Entanglement_2015,
  title = {Entanglement control in quantum networks by quantum-coherent time-delayed feedback},
  author = {Hein, Sven M. and Schulze, Franz and Carmele, Alexander and Knorr, Andreas},
  journal = {Phys. Rev. A},
  volume = {91},
  issue = {5},
  pages = {052321},
  numpages = {6},
  year = {2015},
  month = {May},
  publisher = {American Physical Society},
  doi = {10.1103/PhysRevA.91.052321},
  url = {https://link.aps.org/doi/10.1103/PhysRevA.91.052321}
}

@misc{zhang_Embedding_2022,
  title = {Embedding of {{Time-Delayed Quantum Feedback}} in a {{Nonreciprocal Array}}},
  author = {Zhang, Xin H. H. and Klapp, S. H. L. and Metelmann, A.},
  year = 2022,
  month = apr,
  number = {arXiv:2204.02367},
  eprint = {2204.02367},
  primaryclass = {quant-ph},
  publisher = {arXiv},
  doi = {10.48550/arXiv.2204.02367},
  urldate = {2026-08-25},
  archiveprefix = {arXiv},
}

@article{pichler_photonic_2016,
	title = {{Photonic Circuits with Time Delays and Quantum Feedback}},
	volume = {116},
	issn = {0031-9007, 1079-7114},
	url = {https://link.aps.org/doi/10.1103/PhysRevLett.116.093601},
	doi = {10.1103/PhysRevLett.116.093601},   
	number = {9},
	urldate = {2022-05-12},
	journal = {Phys. Rev. Lett.},
	author = {Pichler, Hannes and Zoller, Peter},
	month = mar,
	year = {2016},
	pages = {093601},
}

@article{Crowder_Quantum_2020,
  title = {{Quantum Trajectory Theory of Few-Photon Cavity-QED Systems with a Time-Delayed Coherent Feedback}},
  author = {Crowder, Gavin and Carmichael, Howard and Hughes, Stephen},
  journal = {Phys. Rev. A},
  volume = {101},
  issue = {2},
  pages = {023807},
  numpages = {13},
  year = {2020},
  month = {Feb},
  publisher = {American Physical Society},
  doi = {10.1103/PhysRevA.101.023807},
  url = {https://link.aps.org/doi/10.1103/PhysRevA.101.023807}
}

@book{gardiner_physics_2015,
	title = {{The Physics of Quantum-Optical Devices}},
	year={2015}, 
	author = {Gardiner, Crispin and Zoller, Peter},
    publisher = {Imperial College Press}
}

@article{grimsmo_time-delayed_2015,
	title = {{Time-Delayed Quantum Feedback Control}},
	volume = {115},
	issn = {0031-9007, 1079-7114},
	url = {https://link.aps.org/doi/10.1103/PhysRevLett.115.060402},
	doi = {10.1103/PhysRevLett.115.060402},
	number = {6},
	urldate = {2022-05-19},
	journal = {Phys. Rev. Lett.},
	author = {Grimsmo, Arne L.},
	month = aug,
	year = {2015},
	pages = {060402},
}

@article{barkemeyer_Heisenberg_2022,
  title = {{Heisenberg Treatment of Multiphoton Pulses in Waveguide QED with Time-Delayed Feedback}},
  author = {Barkemeyer, Kisa and Knorr, Andreas and Carmele, Alexander},
  journal = {Phys. Rev. A},
  volume = {106},
  issue = {2},
  pages = {023708},
  numpages = {10},
  year = {2022},
  month = {Aug},
  publisher = {American Physical Society},
  doi = {10.1103/PhysRevA.106.023708},
  url = {https://link.aps.org/doi/10.1103/PhysRevA.106.023708}
}

@article{ferreira_Deterministic_2024,
  title = {Deterministic Generation of Multidimensional Photonic Cluster States with a Single Quantum Emitter},
  author = {Ferreira, Vinicius S. and Kim, Gihwan and Butler, Andreas and Pichler, Hannes and Painter, Oskar},
  year = 2024,
  month = may,
  journal = {Nature Physics},
  volume = {20},
  number = {5},
  pages = {865--870},
  publisher = {Nature Publishing Group},
  issn = {1745-2481},
  doi = {10.1038/s41567-024-02408-0},
  urldate = {2026-08-31},
  copyright = {2024 The Author(s), under exclusive licence to Springer Nature Limited},
  langid = {english},
}

@Article{hoi_Probing_2015,
author={Hoi, I.-C.
and Kockum, A. F.
and Tornberg, L.
and Pourkabirian, A.
and Johansson, G.
and Delsing, P.
and Wilson, C. M.},
title={{Probing the Quantum Vacuum with an Artificial Atom in Front of a Mirror}},
journal={Nature Physics},
year={2015},
month={Dec},
day={01},
volume={11},
number={12},
pages={1045-1049},
issn={1745-2481},
doi={10.1038/nphys3484},
url={https://doi.org/10.1038/nphys3484}
}

@article{crowder_Improving_2024,
  title = {Improving on-demand single-photon-source coherence and indistinguishability through a time-delayed coherent feedback},
  author = {Crowder, Gavin and Ramunno, Lora and Hughes, Stephen},
  journal = {Phys. Rev. A},
  volume = {110},
  issue = {3},
  pages = {L031703},
  numpages = {7},
  year = {2024},
  month = {Sep},
  publisher = {American Physical Society},
  doi = {10.1103/PhysRevA.110.L031703},
  url = {https://link.aps.org/doi/10.1103/PhysRevA.110.L031703}
}

@article{whalen_open_2017,
	title = {{Open Quantum Systems with Delayed Coherent Feedback}},
	volume = {2},
	issn = {2058-9565},
	url = {https://iopscience.iop.org/article/10.1088/2058-9565/aa8331},
	doi = {10.1088/2058-9565/aa8331},   
	number = {4},
	urldate = {2022-09-28},
	journal = {Quantum Sci. Technol.},
	author = {Whalen, S J and Grimsmo, A L and Carmichael, H J},
	month = dec,
	year = {2017},
	pages = {044008},
}

@Article{Busnaina2024,
author={Busnaina, Jamal H.
and Shi, Zheng
and McDonald, Alexander
and Dubyna, Dmytro
and Nsanzineza, Ibrahim
and Hung, Jimmy S. C.
and Chang, C. W. Sandbo
and Clerk, Aashish A.
and Wilson, Christopher M.},
title={Quantum simulation of the bosonic Kitaev chain},
journal={Nature Communications},
year={2024},
month={Apr},
day={09},
volume={15},
number={1},
pages={3065},
issn={2041-1723},
doi={10.1038/s41467-024-47186-8},
url={https://doi.org/10.1038/s41467-024-47186-8}
}

@article{Xiao2021,
  title = {Observation of Non-Bloch Parity-Time Symmetry and Exceptional Points},
  author = {Xiao, Lei and Deng, Tianshu and Wang, Kunkun and Wang, Zhong and Yi, Wei and Xue, Peng},
  journal = {Phys. Rev. Lett.},
  volume = {126},
  issue = {23},
  pages = {230402},
  numpages = {6},
  year = {2021},
  month = {Jun},
  publisher = {American Physical Society},
  doi = {10.1103/PhysRevLett.126.230402},
  url = {https://link.aps.org/doi/10.1103/PhysRevLett.126.230402}
}

@article{Ghatak2020,
author = {Ananya Ghatak  and Martin Brandenbourger  and Jasper van Wezel  and Corentin Coulais },
title = {Observation of non-Hermitian topology and its bulk–edge correspondence in an active mechanical metamaterial},
journal = {Proceedings of the National Academy of Sciences},
volume = {117},
number = {47},
pages = {29561-29568},
year = {2020},
doi = {10.1073/pnas.2010580117},
URL = {https://www.pnas.org/doi/abs/10.1073/pnas.2010580117},
eprint = {https://www.pnas.org/doi/pdf/10.1073/pnas.2010580117},
}

@article{Weidemann2020,
author = {Sebastian Weidemann  and Mark Kremer  and Tobias Helbig  and Tobias Hofmann  and Alexander Stegmaier  and Martin Greiter  and Ronny Thomale  and Alexander Szameit },
title = {Topological funneling of light},
journal = {Science},
volume = {368},
number = {6488},
pages = {311-314},
year = {2020},
doi = {10.1126/science.aaz8727},
URL = {https://www.science.org/doi/abs/10.1126/science.aaz8727},
eprint = {https://www.science.org/doi/pdf/10.1126/science.aaz8727},
}

@article{Wang2021,
author = {Kai Wang  and Avik Dutt  and Ki Youl Yang  and Casey C. Wojcik  and Jelena Vučković  and Shanhui Fan },
title = {Generating arbitrary topological windings of a non-Hermitian band},
journal = {Science},
volume = {371},
number = {6535},
pages = {1240-1245},
year = {2021},
doi = {10.1126/science.abf6568},
URL = {https://www.science.org/doi/abs/10.1126/science.abf6568},
eprint = {https://www.science.org/doi/pdf/10.1126/science.abf6568},
}

@Article{Wang2021_2,
author={Wang, Kai
and Dutt, Avik
and Wojcik, Charles C.
and Fan, Shanhui},
title={Topological complex-energy braiding of non-Hermitian bands},
journal={Nature},
year={2021},
month={Oct},
day={01},
volume={598},
number={7879},
pages={59-64},
issn={1476-4687},
doi={10.1038/s41586-021-03848-x},
url={https://doi.org/10.1038/s41586-021-03848-x}
}

@article{Yao2018,
  title = {Edge States and Topological Invariants of Non-Hermitian Systems},
  author = {Yao, Shunyu and Wang, Zhong},
  journal = {Phys. Rev. Lett.},
  volume = {121},
  issue = {8},
  pages = {086803},
  numpages = {8},
  year = {2018},
  month = {Aug},
  publisher = {American Physical Society},
  doi = {10.1103/PhysRevLett.121.086803},
  url = {https://link.aps.org/doi/10.1103/PhysRevLett.121.086803}
}

@article{Okuma2020,
  title = {Topological Origin of Non-Hermitian Skin Effects},
  author = {Okuma, Nobuyuki and Kawabata, Kohei and Shiozaki, Ken and Sato, Masatoshi},
  journal = {Phys. Rev. Lett.},
  volume = {124},
  issue = {8},
  pages = {086801},
  numpages = {7},
  year = {2020},
  month = {Feb},
  publisher = {American Physical Society},
  doi = {10.1103/PhysRevLett.124.086801},
  url = {https://link.aps.org/doi/10.1103/PhysRevLett.124.086801}
}

@article{McDonald2022,
  title = {Nonequilibrium stationary states of quantum non-Hermitian lattice models},
  author = {McDonald, A. and Hanai, R. and Clerk, A. A.},
  journal = {Phys. Rev. B},
  volume = {105},
  issue = {6},
  pages = {064302},
  numpages = {19},
  year = {2022},
  month = {Feb},
  publisher = {American Physical Society},
  doi = {10.1103/PhysRevB.105.064302},
  url = {https://link.aps.org/doi/10.1103/PhysRevB.105.064302}
}

@article{Joshi2023,
  title = {Resonance Fluorescence of a Chiral Artificial Atom},
  author = {Joshi, Chaitali and Yang, Frank and Mirhosseini, Mohammad},
  journal = {Phys. Rev. X},
  volume = {13},
  issue = {2},
  pages = {021039},
  numpages = {27},
  year = {2023},
  month = {Jun},
  publisher = {American Physical Society},
  doi = {10.1103/PhysRevX.13.021039},
  url = {https://link.aps.org/doi/10.1103/PhysRevX.13.021039}
}

@Article{Kannan2023,
author={Kannan, Bharath
and Almanakly, Aziza
and Sung, Youngkyu
and Di Paolo, Agustin
and Rower, David A.
and Braum{\"u}ller, Jochen
and Melville, Alexander
and Niedzielski, Bethany M.
and Karamlou, Amir
and Serniak, Kyle
and Veps{\"a}l{\"a}inen, Antti
and Schwartz, Mollie E.
and Yoder, Jonilyn L.
and Winik, Roni
and Wang, Joel I-Jan
and Orlando, Terry P.
and Gustavsson, Simon
and Grover, Jeffrey A.
and Oliver, William D.},
title={On-demand directional microwave photon emission using waveguide quantum electrodynamics},
journal={Nature Physics},
year={2023},
month={Mar},
day={01},
volume={19},
number={3},
pages={394-400},
issn={1745-2481},
doi={10.1038/s41567-022-01869-5},
url={https://doi.org/10.1038/s41567-022-01869-5}
}

@article{irfan_Autonomous_2026,
  title = {Autonomous {{Stabilization}} of {{Remote Entanglement}} in a {{Cascaded Quantum Network}}},
  author = {Irfan, Abdullah and Singirikonda, Kaushik and Yao, Mingxing and Lingenfelter, Andrew and Mollenhauer, Michael and Cao, Xi and Clerk, Aashish A. and Pfaff, Wolfgang},
  year = 2026,
  month = jul,
  journal = {Physical Review X},
  volume = {16},
  number = {3},
  pages = {031004},
  publisher = {American Physical Society},
  doi = {10.1103/z6zz-vw5q},
  urldate = {2026-07-13},
}

@article{irfan2024,
  title = {Loss resilience of driven-dissipative remote entanglement in chiral waveguide quantum electrodynamics},
  author = {Irfan, Abdullah and Yao, Mingxing and Lingenfelter, Andrew and Cao, Xi and Clerk, Aashish A. and Pfaff, Wolfgang},
  journal = {Phys. Rev. Res.},
  volume = {6},
  issue = {3},
  pages = {033212},
  numpages = {11},
  year = {2024},
  month = {Aug},
  publisher = {American Physical Society},
  doi = {10.1103/PhysRevResearch.6.033212},
  url = {https://link.aps.org/doi/10.1103/PhysRevResearch.6.033212}
}

@article{Ashida2020,
author = {Yuto Ashida and Zongping Gong and Masahito Ueda},
title = {Non-Hermitian physics},
journal = {Advances in Physics},
volume = {69},
number = {3},
pages = {249--435},
year = {2020},
publisher = {Taylor \& Francis},
doi = {10.1080/00018732.2021.1876991},
URL = {https://doi.org/10.1080/00018732.2021.1876991},
eprint = {https://doi.org/10.1080/00018732.2021.1876991}
}

@article{Bergholtz2021,
  title={Exceptional topology of non-Hermitian systems},
  author={Bergholtz, Emil J and Budich, Jan Carl and Kunst, Flore K},
  journal={Reviews of Modern Physics},
  volume={93},
  number={1},
  pages={015005},
  year={2021},
  publisher={American Physical Society},
  doi = {10.1103/RevModPhys.93.015005},
  URL = {https://doi.org/10.1103/RevModPhys.93.015005}
}

@article{Haga2021,
  title = {Liouvillian Skin Effect: Slowing Down of Relaxation Processes without Gap Closing},
  author = {Haga, Taiki and Nakagawa, Masaya and Hamazaki, Ryusuke and Ueda, Masahito},
  journal = {Phys. Rev. Lett.},
  volume = {127},
  issue = {7},
  pages = {070402},
  numpages = {7},
  year = {2021},
  month = {Aug},
  publisher = {American Physical Society},
  doi = {10.1103/PhysRevLett.127.070402},
  url = {https://link.aps.org/doi/10.1103/PhysRevLett.127.070402}
}

@article{lee_Anomalously_2023,
  title = {Anomalously Large Relaxation Times in Dissipative Lattice Models beyond the Non-{{Hermitian}} Skin Effect},
  author = {Lee, Gideon and McDonald, Alexander and Clerk, Aashish},
  year = 2023,
  month = aug,
  journal = {Physical Review B},
  volume = {108},
  number = {6},
  pages = {064311},
  publisher = {American Physical Society},
  doi = {10.1103/PhysRevB.108.064311},
  urldate = {2024-04-03},
}

@article{Begg2024,
  title = {Quantum Criticality in Open Quantum Spin Chains with Nonreciprocity},
  author = {Begg, Samuel E. and Hanai, Ryo},
  journal = {Phys. Rev. Lett.},
  volume = {132},
  issue = {12},
  pages = {120401},
  numpages = {7},
  year = {2024},
  month = {Mar},
  publisher = {American Physical Society},
  doi = {10.1103/PhysRevLett.132.120401},
  url = {https://link.aps.org/doi/10.1103/PhysRevLett.132.120401}
}

@article{Yang2023,
  title = {Efficient Information Retrieval for Sensing via Continuous Measurement},
  author = {Yang, Dayou and Huelga, Susana F. and Plenio, Martin B.},
  journal = {Phys. Rev. X},
  volume = {13},
  issue = {3},
  pages = {031012},
  numpages = {27},
  year = {2023},
  month = {Jul},
  publisher = {American Physical Society},
  doi = {10.1103/PhysRevX.13.031012},
  url = {https://link.aps.org/doi/10.1103/PhysRevX.13.031012}
}

@article{trefethen2005,
  title={Spectra and pseudospectra. Princeton Univ},
  author={Trefethen, Lloyd N and Embree, Mark},
  journal={Press, Princeton, NJ},
  year={2005}
}

@article{ferreira_Collapse_2021,
  title = {Collapse and {{Revival}} of an {{Artificial Atom Coupled}} to a {{Structured Photonic Reservoir}}},
  author = {Ferreira, Vinicius S. and Banker, Jash and Sipahigil, Alp and Matheny, Matthew H. and Keller, Andrew J. and Kim, Eunjong and Mirhosseini, Mohammad and Painter, Oskar},
  year = 2021,
  month = dec,
  journal = {Physical Review X},
  volume = {11},
  number = {4},
  pages = {041043},
  publisher = {American Physical Society},
  doi = {10.1103/PhysRevX.11.041043},
  urldate = {2025-12-17},
}

@article{mirhosseini_Superconducting_2018,
  title = {Superconducting Metamaterials for Waveguide Quantum Electrodynamics},
  author = {Mirhosseini, Mohammad and Kim, Eunjong and Ferreira, Vinicius S. and Kalaee, Mahmoud and Sipahigil, Alp and Keller, Andrew J. and Painter, Oskar},
  year = 2018,
  month = sep,
  journal = {Nature Communications},
  volume = {9},
  number = {1},
  pages = {3706},
  publisher = {Nature Publishing Group},
  issn = {2041-1723},
  doi = {10.1038/s41467-018-06142-z},
  urldate = {2025-12-17},
}

@Article{Jordan1928,
author={Jordan, P.
and Wigner, E.},
title={{\"U}ber das Paulische {\"A}quivalenzverbot},
journal={Zeitschrift f{\"u}r Physik},
year={1928},
month={Sep},
day={01},
volume={47},
number={9},
pages={631-651},
issn={0044-3328},
doi={10.1007/BF01331938},
url={https://doi.org/10.1007/BF01331938}
}

@article{Pocklington2025,
  title = {Efficient Simulation of Nontrivial Dissipative Spin Chains via Stochastic Unraveling},
  author = {Pocklington, Andrew and Clerk, Aashish A.},
  journal = {PRX Quantum},
  volume = {6},
  issue = {3},
  pages = {030349},
  numpages = {29},
  year = {2025},
  month = {Sep},
  publisher = {American Physical Society},
  doi = {10.1103/vptq-xy6h},
  url = {https://link.aps.org/doi/10.1103/vptq-xy6h}
}

@article{Combes2017,
author = {Joshua Combes and Joseph Kerckhoff and Mohan Sarovar},
title = {The SLH framework for modeling quantum input-output networks},
journal = {Advances in Physics: X},
volume = {2},
number = {3},
pages = {784--888},
year = {2017},
publisher = {Taylor \& Francis},
doi = {10.1080/23746149.2017.1343097},
URL = {https://doi.org/10.1080/23746149.2017.1343097},
eprint ={https://doi.org/10.1080/23746149.2017.1343097}
}

@article{Torres2014,
  title = {Closed-form solution of Lindblad master equations without gain},
  author = {Torres, Juan Mauricio},
  journal = {Phys. Rev. A},
  volume = {89},
  issue = {5},
  pages = {052133},
  numpages = {9},
  year = {2014},
  month = {May},
  publisher = {American Physical Society},
  doi = {10.1103/PhysRevA.89.052133},
  url = {https://link.aps.org/doi/10.1103/PhysRevA.89.052133}
}

@article{liang2022dynamic,
  title = {Dynamic Signatures of Non-Hermitian Skin Effect and Topology in Ultracold Atoms},
  author = {Liang, Qian and Xie, Dizhou and Dong, Zhaoli and Li, Haowei and Li, Hang and Gadway, Bryce and Yi, Wei and Yan, Bo},
  journal = {Phys. Rev. Lett.},
  volume = {129},
  issue = {7},
  pages = {070401},
  numpages = {6},
  year = {2022},
  month = {Aug},
  publisher = {American Physical Society},
  doi = {10.1103/PhysRevLett.129.070401}
}

@book{pozar_microwave_2012,
  title = {Microwave Engineering},
  author = {Pozar, David M.},
  year = 2012,
  month = apr,
  edition = {4},
  publisher = {John Wiley and Sons, Inc.},
  address = {Hoboken, NJ},
  isbn = {978-0-470-63155-3},
}

@article{Gammelmark2013past,
  title = {Past Quantum States of a Monitored System},
  author = {Gammelmark, S\o{}ren and Julsgaard, Brian and M\o{}lmer, Klaus},
  journal = {Phys. Rev. Lett.},
  volume = {111},
  issue = {16},
  pages = {160401},
  numpages = {5},
  year = {2013},
  month = {Oct},
  publisher = {American Physical Society},
  doi = {10.1103/PhysRevLett.111.160401}
}

@article{bao2020spin,
  title={Spin squeezing of 1011 atoms by prediction and retrodiction measurements},
  author={Bao, Han and Duan, Junlei and Jin, Shenchao and Lu, Xingda and Li, Pengxiong and Qu, Weizhi and Wang, Mingfeng and Novikova, Irina and Mikhailov, Eugeniy E and Zhao, Kai-Feng and others},
  journal={Nature},
  volume={581},
  number={7807},
  pages={159--163},
  year={2020},
  publisher={Nature Publishing Group UK London},
  doi = {10.1038/s41586-020-2243-7}
}

@article{astafiev2010resonance,
  title={Resonance fluorescence of a single artificial atom},
  author={Astafiev, O and Zagoskin, Alexandre M and Abdumalikov Jr, AA and Pashkin, Yu A and Yamamoto, T and Inomata, K and Nakamura, Y and Tsai, Jaw Shen},
  journal={Science},
  volume={327},
  number={5967},
  pages={840--843},
  year={2010},
  publisher={American Association for the Advancement of Science},
  doi = {10.1126/science.1181918}
}

@article{arcari2014near,
  title = {Near-Unity Coupling Efficiency of a Quantum Emitter to a Photonic Crystal Waveguide},
  author = {Arcari, M. and S\"ollner, I. and Javadi, A. and Lindskov Hansen, S. and Mahmoodian, S. and Liu, J. and Thyrrestrup, H. and Lee, E. H. and Song, J. D. and Stobbe, S. and Lodahl, P.},
  journal = {Phys. Rev. Lett.},
  volume = {113},
  issue = {9},
  pages = {093603},
  numpages = {5},
  year = {2014},
  month = {Aug},
  publisher = {American Physical Society},
  doi = {10.1103/PhysRevLett.113.093603},
  url = {https://link.aps.org/doi/10.1103/PhysRevLett.113.093603}
}

@article{weidemann2022topological,
  title={Topological triple phase transition in non-Hermitian Floquet quasicrystals},
  author={Weidemann, Sebastian and Kremer, Mark and Longhi, Stefano and Szameit, Alexander},
  journal={Nature},
  volume={601},
  number={7893},
  pages={354--359},
  year={2022},
  publisher={Nature Publishing Group UK London},
  doi={10.1038/s41586-021-04253-0}
}

@article{PhysRevLett.131.033606,
  title = {Independent Electrical Control of Two Quantum Dots Coupled through a Photonic-Crystal Waveguide},
  author = {Chu, Xiao-Liu and Papon, Camille and Bart, Nikolai and Wieck, Andreas D. and Ludwig, Arne and Midolo, Leonardo and Rotenberg, Nir and Lodahl, Peter},
  journal = {Phys. Rev. Lett.},
  volume = {131},
  issue = {3},
  pages = {033606},
  numpages = {6},
  year = {2023},
  month = {Jul},
  publisher = {American Physical Society},
  doi = {10.1103/PhysRevLett.131.033606}
}

@article{sollner2015deterministic,
  title={Deterministic photon--emitter coupling in chiral photonic circuits},
  author={S{\"o}llner, Immo and Mahmoodian, Sahand and Hansen, Sofie Lindskov and Midolo, Leonardo and Javadi, Alisa and Kir{\v{s}}ansk{\.e}, Gabija and Pregnolato, Tommaso and El-Ella, Haitham and Lee, Eun Hye and Song, Jin Dong and others},
  journal={Nature nanotechnology},
  volume={10},
  number={9},
  pages={775--778},
  year={2015},
  publisher={Nature Publishing Group UK London},
  doi={10.1038/nnano.2015.159}
}

@article{almanakly_Deterministic_2025,
  title = {Deterministic Remote Entanglement Using a Chiral Quantum Interconnect},
  author = {Almanakly, Aziza and Yankelevich, Beatriz and Hays, Max and Kannan, Bharath and Assouly, R{\'e}ouven and Greene, Alex and Gingras, Michael and Niedzielski, Bethany M. and Stickler, Hannah and Schwartz, Mollie E. and Serniak, Kyle and Wang, Joel {\^I}-j and Orlando, Terry P. and Gustavsson, Simon and Grover, Jeffrey A. and Oliver, William D.},
  year = 2025,
  month = mar,
  journal = {Nature Physics},
  pages = {1--6},
  publisher = {Nature Publishing Group},
  issn = {1745-2481},
  doi = {10.1038/s41567-025-02811-1},
  urldate = {2025-04-10},
  copyright = {2025 The Author(s), under exclusive licence to Springer Nature Limited},
  langid = {english},
}

@article{agusti_Autonomous_2023,
  title = {Autonomous {{Distribution}} of {{Programmable Multiqubit Entanglement}} in a {{Dual-Rail Quantum Network}}},
  author = {Agust{\'i}, J. and Zhang, X. H. H. and Minoguchi, Y. and Rabl, P.},
  year = 2023,
  month = dec,
  journal = {Physical Review Letters},
  volume = {131},
  number = {25},
  pages = {250801},
  publisher = {American Physical Society},
  doi = {10.1103/PhysRevLett.131.250801},
  urldate = {2024-03-22},
}

@article{zou_Optimal_2026,
title = {Optimal Generation of Loss-tolerant Photonic Tree Cluster States via Time-Delayed Feedback},
journal = {Chin. Phys. Lett.},
volume = {43},
number = {9},
pages = {090601},
year = {2026},
issn = {},
doi = {10.1088/0256-307X/43/9/090601},	
url = {http://cpl.iphy.ac.cn/en/article/doi/10.1088/0256-307X/43/9/090601},
author = {Jia-Jin Zou and Jian-Wei Qin and Ze-Liang Xiang}
}

@inproceedings{regidor_qwavemps_2026,
author = {Sofia Arranz Regidor and Matthew Kozma and Stephen Hughes},
title = {{QwaveMPS: an efficient matrix-product-states Python package for simulating non-Markovian waveguide-QED systems}},
volume = {PC14143},
booktitle = {Photonics for Quantum 2026},
editor = {Michael Reimer and Nir Rotenberg and Lindsay LeBlanc},
organization = {International Society for Optics and Photonics},
publisher = {SPIE},
pages = {PC141430K},
year = {2026},
doi = {10.1117/12.3109239},
URL = {https://doi.org/10.1117/12.3109239}
}

@misc{daraban_Universal_2026,
  title = {Universal Dynamics from a Single-Particle Dark State},
  author = {Daraban, Ruben and {Safavi-Naini}, Arghavan and Schachenmayer, Johannes and Maghrebi, Mohammad},
  year = 2026,
  month = may,
  number = {arXiv:2605.16494},
  eprint = {2605.16494},
  primaryclass = {cond-mat.quant-gas},
  publisher = {arXiv},
  doi = {10.48550/arXiv.2605.16494},
  urldate = {2026-07-13},
  archiveprefix = {arXiv}
}

\onecolumngrid
\begin{center} \textbf{End Matter}\end{center}
\twocolumngrid
We present here a simple proof that the delayed coherent feedback leads to an effective cascaded interaction between single QEs at different times. To this end, we first notice that the atom-photon interaction Hamiltonian can be generally formulated as \cite{Vodenkova2024}
\begin{equation}
H_{\text{int}}(t)=\mathbbm{i}\sqrt{\kappa}\left([\sigma^\rightarrow b^\dagger(t+\tau)+ \sigma^\leftarrow b^\dagger (t)e^{\mathbbm{i}\phi}]-\rm h.c.\right) \label{eq:Hamiltonian}
\end{equation}
in the frame rotating with atomic transition frequency $\omega_0$ and the interaction picture with
respect to the waveguide bath Hamiltonian. Here, $b(t)$ denotes the quantum noise operator for the waveguide field, satisfying the bosonic commutation relation $[b(t),b^\dag (t')]=\delta(t-t^\prime)$ \cite{gardiner_physics_2015}, with $\phi=\pi-\omega_0\tau$ the phase of a photon acquired after the round trip. $\sigma^{\rightleftarrows}$ are spin operators of the QE associated with the coupling to the right- and left-propagating fields, respectively. For the symmetric coupling considered in the main text, they are spin-lowering operators $\sigma^\rightarrow =\sigma^\leftarrow=\sigma^-$, while asymmetric spin operators can describe a chiral waveguide QED where the atom-photon coupling strengths are directionally dependent \cite{almanakly_Deterministic_2025}.

\begin{figure}[b]
	\centering
	\includegraphics[width=\linewidth]{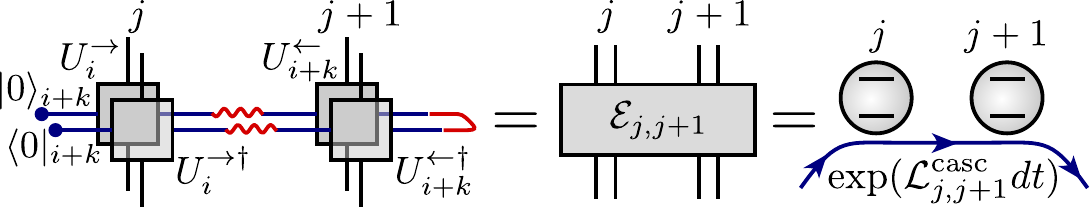}
	\caption{Tensor network illustration of the effective cascaded interactions. Here, the tensor contraction between unitary operators at times $t_i=idt$ and $t_{i+k}=t_i+\tau$ yields an effective propagator $\exp{(\mathcal{L}_{j,j+1}^\mathrm{casc}dt)}$ for the QE.}
	\label{fig:tensor}
\end{figure}

Equation \eqref{eq:Hamiltonian} implies that the QE interacts at time $t$ with the right-going field $b(t+\tau)$, which then returns as a delayed left-going field and interacts with the same QE at time $t+\tau$. Therefore, it is expected that an effective coupling between QEs at time $t$ and $t+\tau$ can be established by tracing-out the bosonic field $b(t+\tau)$. To get a quantitative description, we discretize the total evolution time in small steps $dt$, such that $t_i=t_0+i dt$ and $\tau=k dt$ with both $i$ and $k$ integers. Introducing Ito increment operators for each time-bin mode $\Delta B_i=\int_{t_i}^{t_i+dt}b(t')dt'$ which obeys
 $[\Delta B_i,\Delta B^\dagger_j]= dt\delta_{ij}$, the full quantum state $\ket{\psi_n}$ at time $t_n=ndt$ evolving from an initial state $\ket{\psi_0}$ takes the form
 \begin{equation}
	 \ket{\psi_n} = U_{n-1}^{\leftarrow}U_{n-1}^{\rightarrow}\cdots U_{i}^{\leftarrow}U_{i}^{\rightarrow}\cdots U_0^\leftarrow U_0^\rightarrow\ket{\psi_0}, \label{eq:evolution}
	 \end{equation}
 where $U_i^\rightarrow$ and $U_i^\leftarrow$ describe the scattering induced by the right- and the left-going field, respectively:
 \begin{align}
	 U_i^\rightarrow &=\exp(\sqrt{\kappa}\sigma^\rightarrow\Delta B_{i+k}^\dag -\rm h.c.), \nonumber \\ 
	 U_{i}^\leftarrow &=\exp(\sqrt{\kappa}e^{\mathbbm{i}\phi}\sigma^\leftarrow\Delta B_{i}^\dag -\rm h.c.).
	 \end{align}
One way to interpret Eq.~\eqref{eq:evolution} is to view each unitary operator as a tensor and their multiplications as tensor contractions over photonic and atomic degrees of freedom. Noting that the time-bin mode $\Delta B_{i+k}$ appears twice ($U_i^\rightarrow$ and $U_{i+k}^\leftarrow$), we first perform tensor contraction over this photonic mode while leaving the QE-dimension free, as illustrated in Fig.~\ref{fig:tensor}. This yields an effective propagator between two replica QEs $j=\left\lceil {i}/{k} \right\rceil$ and $j+1$:
\begin{equation}
\mathcal{E}_{j,j+1}=\mathrm{Tr}_{i+k}\left[U_{i+k}^\leftarrow U_i^\rightarrow \ket{0}\bra{0}_{i+k}U_i^{\rightarrow\dagger} U_{i+k}^{\leftarrow\dagger}\right],
\end{equation}
where the mode labeled ${i+k}$ is assumed to be initially in a vacuum state $\ket{0}_{i+k}$. Up to the first order in $dt$, the propagator takes the form $\mathcal{E}_{j,j+1}=\exp(\mathcal{L}_{j,j+1}^\mathrm{casc}dt)+\mathcal{O}(dt^2)$, where the effective Liouvillian $\mathcal{L}_{j,j+1}^\mathrm{casc}$ corresponds to a cascaded interaction between $j$-th and $(j+1)$-th replica systems, i.e.,
\begin{equation}
\mathcal{L}^{\rm casc}_{j,j+1}(\rho)=-\mathbbm{i}\left[H_{j,j+1}^{\rm casc},\rho\right]+\kappa \mathcal{D}[\sigma_j^\rightarrow+e^{\mathbbm{i}\phi}\sigma_{j+1}^\leftarrow]\rho
\end{equation}
with $H_{j,j+1}^{\rm casc}=\mathbbm{i}\kappa(e^{\mathbbm{i}\phi} \sigma_j^{\rightarrow\dag} \sigma_{j+1}^\leftarrow-\text{h.c.})/2$ and $\mathcal{D}[\sigma]\rho=\sigma \rho \sigma^\dag-(\sigma^\dag \sigma \rho + \rho \sigma^\dag \sigma)/2$. The above discussion reveals the equivalence between the cascaded interaction and the coherent delayed feedback, as introduced in Ref.~\cite{grimsmo_time-delayed_2015} and used in Ref.~\cite{Vodenkova2024} to compute the QE dynamics. The derivations presented here can be generalized to the case with multiple feedback loops \cite{SMarxiv}, which lead to a general cascaded master equation Eqs.~\eqref{eq:master} and \eqref{eq:general}.

\end{document}

\title{Supplementary Materials for: Efficient Simulation of Nonreciprocal Many-body Physics via Quantum Feedback}

\author{K. Vodenkova}
\affiliation{Institute for Theoretical Physics, University of Innsbruck, 6020 Innsbruck, Austria \\ and Institute for Quantum Optics and Quantum Information of the Austrian Academy of Sciences, \\ 6020 Innsbruck, Austria}

\author{A. Pocklington}
\affiliation{Pritzker School of Molecular Engineering, University of Chicago, Chicago, IL 60637, USA}
\affiliation{Department of Physics, University of Chicago, Chicago, IL 60637, USA}

\author{F. Yang}
\affiliation{School of Physics and Zhejiang Key Laboratory of Micro-nano Quantum Chips \\ and Quantum Control, Zhejiang University, Hangzhou 310027, China}

\author{A. Lingenfelter}
\affiliation{Institute for Theoretical Physics, University of Innsbruck, 6020 Innsbruck, Austria \\ and Institute for Quantum Optics and Quantum Information of the Austrian Academy of Sciences, \\ 6020 Innsbruck, Austria}

\author{A. A. Clerk}
\affiliation{Pritzker School of Molecular Engineering, University of Chicago, Chicago, IL 60637, USA}

\author{H. Pichler}
\affiliation{Institute for Theoretical Physics, University of Innsbruck, 6020 Innsbruck, Austria \\ and Institute for Quantum Optics and Quantum Information of the Austrian Academy of Sciences, \\ 6020 Innsbruck, Austria}

\date{\today}

\maketitle

\tableofcontents

\section{Effective spin model induced by a general feedback network}
\subsection{Derivation of the effective Liouvillian}
As described in the main text, under multiple feedback loops, the dynamics for an $N$-site effective spin chain is governed by a many-body nonreciprocal model
\begin{align}
\partial_t \rho=-\mathbbm{i}\left(H_{\rm NH}\rho-\rho H_{\rm NH}^\dagger\right)+\sum_{i,j}J_{ij}\sigma_{i}\rho \sigma_{j}^\dagger+\sum_{i,\mu}\mathcal{D}[L_{i}^\mu]\rho, \label{eq:master}
\end{align}
with $i,j=1,2,\cdots,N$ and a non-Hermitian Hamiltonian
\begin{align}
    H_{\rm NH}=\sum_{i} \left(H_{\mathrm{s},i}-{\mathbbm{i}}J_{ii}\sigma^\dagger_{i}\sigma_{i}/2\right)-{\mathbbm{i}}\sum_{i<j}J_{ij}\sigma^\dagger_{j}\sigma_{i}, \label{eq:general}
\end{align}
where the coupling coefficient $J_{ij}$ is determined by the configuration of feedback loops.
Here, we let $\sigma_i$ be an arbitrary spin operator of site $i$, not necessarily the lowering operator $\sigma_i^-$, to demonstrate the generality of the scheme.

For the configuration shown in Fig.~1(b) of the main text, each feedback loop $n$ with a delay time $n\tau$ is enabled by an orthogonal waveguide mode $b_n(t)$, such that the original atom-photon interaction Hamiltonian takes the form
\begin{equation}
    H_{\rm int}(t)=\mathbbm{i}\sum_{n=1}^K\sqrt{\kappa}_n
\left(\left[\sigma b_n^\dagger(t+n\tau)+\sigma b_n^\dagger (t)e^{i\phi_n}\right]-\mathrm{h.c.}\right).\end{equation}
Therefore, we can treat each feedback loop independently based on the same analysis presented in the main text, i.e., each loop will generate a two-site cascaded channel between the $j$-th and the $(j+l)$-th replica QEs. As a result, we can directly obtain a $K$-local coupling coefficient
\begin{equation}
J_{ij} = \left(2\sum_{n=1}^K\kappa_n\right)\delta_{ij} + \sum_{n=1}^K\left(\kappa_n e^{i\phi_n}\delta_{i,j-n}+\kappa_n e^{-i\phi_n}\delta_{i,j+n}\right).
\end{equation}

For the configuration shown in Fig.~1(c) of the main text, assuming the Bragg mirror is perfectly reflecting, the atom-photon interaction Hamiltonian is then given by
\begin{equation}
    H_{\rm int}(t)=\mathbbm{i}
\sqrt{\kappa}\left(\left[ \sigma b^\dagger (t)+c_1^* \sigma b^\dagger(t-\tau)+c_2^*\sigma b^\dagger (t-2\tau)+\cdots\right]-\mathrm{h.c.}\right)=\mathbbm{i}
\sqrt{\kappa}\sum_{n=0}^{\infty}c_n^*\sigma b^\dagger(t-n\tau)-\mathrm{h.c.},\label{eq:testHam}
\end{equation}
where $c_n=1,2,3,\cdots$ are reflection coefficients (see the next subsection), and $c_0=1$ is defined for notational simplicity. In this case, the time-bin mode $\Delta B_q=\int_{t_q}^{t_q+dt}b(t')dt'$ at $t_q=q\tau+ t$ ($0<t<\tau$) induces a collective cascaded channel $L_{q}^\mathrm{casc}=\sum_{i=\max\{1,q\}}^{N} c_{i-q}^*\sigma_i$. Summing over all relevant cascaded channels ($q=N,N-1,\cdots,-\infty$), we can obtain a Lindblad master equation governing the instantaneous evolution for the effective spin chain at time $t$:
\begin{equation}
\partial_t\rho=\sum_{q=-\infty}^N\left(-\mathbbm{i}\left[H_{q}^{\rm casc},\rho\right]+\kappa \mathcal{D}\left[L_{q}^\mathrm{casc}\right]\rho\right) \quad \textrm{with} \quad
H_{q}^{\rm casc}=\mathbbm{i}\frac{\kappa}{2}\sum_{i=\max\{1,q\}}^{N-1}\sum_{i^\prime=i+1}^N\left(c_{i-q}c_{i^\prime-q}^* \sigma_i^\dagger \sigma_{i^\prime}-\text{h.c.}\right).\label{eq:master_full}
\end{equation}
Equation \eqref{eq:master_full} can be further simplified into the standard form Eqs.~\eqref{eq:master} and \eqref{eq:general} with a long-range coupling coefficient
\begin{equation}
J_{ii} = \kappa\sum_{n=0}^\infty|c_n|^2 = 2\kappa,\quad J_{ij} = \kappa\sum_{n=0}^\infty c_n^*c_{n+j-i} = \kappa c_{j-i}\quad \textrm{and}\quad J_{ji}=J_{ij}^* \quad \textrm{for $i< j$},
\end{equation}
where we utilize the energy conservation law $\sum_{n=1}^\infty |c_n|^2=1$ and the causality constraint $\sum_{n=0}^\infty c_n^*c_{n+d} = c_{d}$ required for any all-pass delay line filter such as the perfectly reflecting Bragg mirror we consider.

\subsection{Engineering a long-range interaction tail}
\begin{figure}
    \centering
    \includegraphics[width=0.9\linewidth]{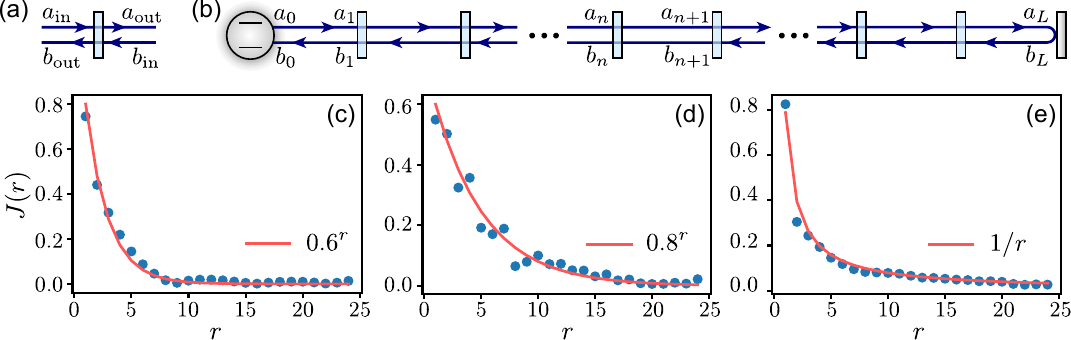}
    \caption{Engineering a long-range interacting cascaded spin chain. (a) Definition of the input and output fields for a partially transmitting mirror. (b) A single atom in front of a Bragg mirror. (c)-(e) show the interaction tails $J(r)=|J_{i,i+r}|\propto 0.6^r$, $0.8^r$, $1/r$, realized by a Bragg mirror of $L=10$ stacks with optimized reflection coefficients $\{r_n,n=1,\cdots, 9\}$ and $r_{10}=-1$.
    }
    \label{fig:fig_s1}
\end{figure}

We now discuss how to engineer a long-range coupling $J_{ij}=\kappa c_{j-i}$. The reflection coefficients $\{c_n\}$ induced by a perfectly reflecting Bragg mirror can be determined using a frequency-domain analysis. First, the scattering relation between the input and output fields for each individual mirror [see Fig.~\ref{fig:fig_s1}(a)] is given by
\begin{equation}
\begin{bmatrix}
a_\mathrm{out}\\ 
b_\mathrm{out}
\end{bmatrix}=\begin{bmatrix}
t&r\\ 
-r^*&t
\end{bmatrix}
\begin{bmatrix}
a_\mathrm{in}\\ 
b_\mathrm{in}
\end{bmatrix},\label{eq:eqs2}
\end{equation}
where $t^2+|r|^2=1$. Equation \eqref{eq:eqs2} is equivalent to the transfer relation
\begin{equation}
\begin{bmatrix}
a_\mathrm{out}\\ 
b_\mathrm{in}
\end{bmatrix}=\begin{bmatrix}
1/t&r/t\\ 
r^*/t&1/t
\end{bmatrix}
\begin{bmatrix}
a_\mathrm{in}\\ 
b_\mathrm{out}
\end{bmatrix}.
\end{equation}
When there are $L$ stacks of equidistantly distributed mirrors separated by an optical path $\phi/2=\omega\tau/2$ [see Fig.~\ref{fig:fig_s1}(b)], the relation between the $n$-th field and $(n+1)$-th field is then given by
\begin{equation}
\mathbf{X}_{n+1}=\mathbf{T}_n\mathbf{X}_n, \textrm{ where }
\mathbf{X}_n=
\begin{bmatrix}
a_n\\ 
b_n
\end{bmatrix},\ 
\mathbf{T}_n=\mathbf{T}^P\mathbf{T}_n^M \textrm{ with }
\mathbf{T}^P=\begin{bmatrix}
e^{\mathbbm{i}\phi/2}&0\\ 
0&e^{-\mathbbm{i}\phi/2}
\end{bmatrix}\textrm{ and }
\mathbf{T}_n^M=
\begin{bmatrix}
1/t_n&r_n/t_n\\ 
r_n^*/t_n&1/t_n
\end{bmatrix}.
\end{equation}
We can then establish the relation between the field $\mathbf{X}_0$ at the quantum emitter and the field $\mathbf{X}_L$ at the last mirror:
\begin{equation}
\mathbf{X}_{L}=\mathbf{T}\mathbf{X}_0, \textrm{ where }\mathbf{T}=\mathbf{T}_{L-1}\mathbf{T}_{L-2}\cdot\mathbf{T}_{2}\mathbf{T}_{1}\mathbf{T}^P.
\end{equation}
Assuming the last mirror is perfectly reflecting ($r_L=-1$), the total reflection coefficient is then given by
\begin{equation}
R = -\frac{\mathbf{T}_{11}+\mathbf{T}_{21}}{\mathbf{T}_{22}+\mathbf{T}_{12}}=\sum_{n=1}^\infty c_ne^{\mathbbm{i}n\phi},
\end{equation}
from which we can determine the coefficient $c_n$ and consequently, the long-range cascaded interaction $J_{ij} = \kappa c_{j-i}$. As a simple example, we can consider a Fabry-Perot (FP) configuration ($L=2$), which leads to 
\begin{equation}
R = -\frac{r^*+e^{\mathbbm{i}\phi}}{r+e^{-\mathbbm{i}\phi}}= -r^* e^{\mathbbm{i}\phi} - t^2\sum_{n=2}^\infty (-r)^{n-2}e^{\mathbbm{i}n\phi},
\end{equation}
i.e., $c_1=-r^*$ and $c_n=(-1)^{n-1}t^2r^{n-2}\ (n>1)$. This implies that an exponentially decaying interaction tail $|J_{ij}|\propto (1/\sqrt{2})^{|i-j|}$ can be established by choosing $|r|^2=1/2$. However, the FP configuration fails to realize a generic long-range interaction tail of different forms. To overcome such a limitation, we can instead consider a multi-stack configuration with $(L-1)$ tunable complex reflection coefficients $\{r_1,r_2,\cdots,r_{L-1}\}$. By optimizing these reflection coefficients, one can mimic long-range interacting tails of different shapes [see Figs.~\ref{fig:fig_s1}(c)-(e)].

\section{Free Fermion Approximation}

The model considered in the main text is given by
\begin{align}
\hat H &= \frac{\Delta}{2} \sum_{i = 1}^N \hat \sigma_i^z - \frac{i \kappa}{2} \sum_{i = 1}^{N-1} \left(\hat \sigma_i^x \hat \sigma_{i + 1}^x + \hat \sigma_i^y \hat \sigma_{i + 1}^y\right), \\
\partial_t \hat \rho &= -i[\hat H, \hat \rho] + \Gamma \sum_{i = 1}^N \D[\hat \sigma^+_i] \hat \rho + \kappa \sum_{i = 0}^N \D[\hat \sigma_i^- + \hat \sigma^-_{i + 1}] \hat \rho.
\end{align}
where we have defined open boundary conditions $\hat \sigma_0 = \hat \sigma_{N + 1} = 0$. To make analytical progress, we can perform a Jordan-Wigner (JW) transform to free fermions, by making the identification \cite{Jordan1928}:
\begin{align}
    \hat c_i &= \left( \prod_{j = 1}^{i - 1} \hat \sigma_j^z \right) \hat \sigma_i^-,
\end{align}
where $\hat c_i$ are canonical fermionic annihilation operators. The new system then becomes:
\begin{align}
\hat H &= \Delta \sum_{i = 1}^N \hat c_i^\dagger \hat c_i - \frac{i\kappa}{2} \sum_{i = 1}^{N-1} \left(\hat c_i^\dagger \hat c_{i + 1} + \hc \right), \\
\hat L_i &= \sqrt{\kappa} \left( \prod_{j = 1}^{i} (-1)^{\hat c_j^\dagger \hat c_j} \right) (\hat c_i + \hat c_{i + 1} ),\ 
\hat L_{\mathrm{pump},i} = \sqrt{\Gamma} \left( \prod_{j = 1}^{i-1} (-1)^{\hat c_j^\dagger \hat c_j} \right) \hat c_i^\dagger .
\end{align}
This system is formally interacting due to the JW string in the jump operator \cite{Pocklington2025}, but the effective Hamiltonian (which governs the single particle spectrum) is completely quadratic and solvable. This is given by
\begin{align}
\hat H_{\mathrm{eff}} &= \sum_{mn} (H_{\mathrm{eff}})_{mn} \hat c_i^\dagger \hat c_j,  \\
 (H_{\mathrm{eff}})_{mn} &= \left( \Delta -i \left[ \kappa + \frac{\Gamma}{2} \right] \right) \delta_{ij} - i \kappa \delta_{i + 1,j}.
\end{align}
The purpose of this section will be to gain exact and analytic insights from the single particle Green's function (which we will call the ``free-fermion'' model) to attempt to gain intuition for the many-body spin problem.

\subsection{Steady State}

Here, we will compute steady state properties of the free fermion model. All relevant properties are encoded in the single particle Green's function:
\begin{align}
G_{ij}(\omega) &= (\omega - H_{\mathrm{eff}} ) ^{-1}_{ij} = \frac{1}{\chi} \times \left\{ 
\begin{array}{cc}
\left(- \frac{i \kappa}{\chi} \right)^{i - j} & i \geq j \\
0 & i < j
\end{array} \right. , \\
\chi &= \omega - \Delta + i\left(\kappa + \frac{\Gamma}{2}\right),
\end{align}
where here $\chi$ is an effective susceptibility, and $\kappa$ is playing the dual role of a loss and hopping rate. Before performing any calculations, we will first note that there is a relevant length scale in the Green's function: 
\begin{align}
\left.\left(- \frac{\kappa}{\chi} \right)^{i - j}\right|_{\omega = \Delta} &= (-1)^{|i-j|} \exp \left( - |i-j|/\xi \right) ,  \\
\xi^{-1} &= \log \left( 1 + \frac{\Gamma}{2 \kappa} \right) \sim \frac{\Gamma}{2 \kappa}, 
\end{align}
where we have assumed weak pumping $\Gamma \ll \kappa$. Clearly, this system becomes critical as $\Gamma/\kappa \to 0$, as the relevant length scale diverges. For sites $j$ such that $j\xi \ll 1$, we expect different behavior than in normal, gapped Lindbladians, \cite{McDonald2022, Begg2024}. 

Now, we can start to calculate the steady state correlation functions, which are given by: 
\begin{align}
\langle \hat c_i^\dagger \hat c_j \rangle &= 2 \Gamma\int \sum_k G_{ik} G^\dagger_{kj} \dd \omega.
\end{align}
In particular, the local densities are given by
\begin{align}
\langle \hat c_i^\dagger \hat c_i \rangle &= 2 \Gamma \sum_{d = 0}^{i}  \int \frac{ \dd \omega}{2 \pi} \frac{\kappa^{2d}}{|\chi|^{2d +2} } = 2 \Gamma \sum_{d = 0}^{i}  \int \frac{ \dd \omega}{2 \pi} \frac{\kappa^{2d}}{(\omega^2 + (\kappa + \Gamma/2)^2)^{d + 1}} \nonumber  \\
&= \frac{1}{\sqrt{\pi}} \frac{ \Gamma}{\kappa + \Gamma/2} \sum_{d = 0}^{i}  \left( \frac{\kappa}{\kappa + \Gamma/2} \right)^{2 d} \frac{\Gamma(d+1/2)}{\Gamma(d + 1)} = \frac{1}{\sqrt{\pi}} \frac{ \Gamma}{\kappa + \Gamma/2} \sum_{d = 0}^{i}  e^{-2 d/\xi} \frac{\Gamma(d+1/2)}{\Gamma(d + 1)}.
\end{align}
Here, $\Gamma(n)$ is the Gamma function, when evaluated on integers it gives $\Gamma(n) = (n-1)!$. Again taking the limit where $d \ll \xi$ we neglect the exponential to recover the limit
\begin{align}
\frac{1}{\sqrt{\pi}} \frac{ \Gamma}{\kappa + \Gamma/2} \sum_{d = 0}^{i}  e^{-2 d/\xi} \frac{\Gamma(d+1/2)}{\Gamma(d + 1)} &\sim  \frac{2\Gamma}{\sqrt{\pi} \kappa} \frac{\Gamma(i + 3/2)}{\Gamma(i + 1)} \xrightarrow{i \gg 1}  \frac{2\Gamma}{\sqrt{\pi}\kappa} \sqrt{i}.
\end{align}

We can also ask what the waveguide output field looks like. Defining the output field mode $\hat a_j = \hat c_j + \hat c_{j + 1}$,  this is given by \cite{Vodenkova2024}:
\begin{align}
\langle \hat a^\dagger_j \hat a_j \rangle &= \langle \hat c_j^\dagger \hat c_j \rangle + \langle \hat c_{j + 1}^\dagger \hat c_j \rangle +  \langle \hat c_{j}^\dagger \hat c_{j + 1} \rangle + \langle \hat c_{j + 1}^\dagger \hat c_{j + 1} \rangle ,
\end{align}
where we have gauged out the phase from the round trip passage. For the off-diagonal term, we can repeat the calculation to observe that:
\begin{align}
\langle \hat c_i^\dagger \hat c_{i + 1} \rangle &= 2 \Gamma \sum_{d = 0}^{i}  \int \frac{\dd \omega}{2 \pi} \frac{\kappa^{2d}}{|\chi|^{2d +2} } \frac{i \kappa}{\chi^*} = 2i \Gamma \sum_{d = 0}^{i}  \int \frac{\dd \omega}{2 \pi} \frac{\kappa^{2d + 1}(\omega + i(\kappa + \Gamma/2))}{(\omega^2 + (\kappa + \Gamma/2)^2)^{d + 2}} . 
\end{align}
Noting that the term that has an odd number of power of $\omega$ vanishes under the integration, this simplifies to
\begin{align}
\langle \hat c_i^\dagger \hat c_{i + 1} \rangle &= -2 \Gamma \sum_{d = 0}^{i}  \int \frac{\dd \omega}{2 \pi} \frac{\kappa^{2d + 1}(\kappa + \Gamma/2)}{(\omega^2 + (\kappa + \Gamma/2)^2)^{d + 2}} \nonumber  \\
&= \frac{-1}{\sqrt{\pi}} \frac{\Gamma}{\kappa + \Gamma/2} \sum_{d = 0}^{i}  e^{-(2 d + 1)/\xi} \frac{\Gamma(d+3/2)}{\Gamma(d + 2)}.
\end{align}
Taking the same limit $\Gamma/\kappa \to 0$ gives us 
\begin{align}
\langle \hat c_i^\dagger \hat c_{i + 1} \rangle &\sim \frac{-1}{\sqrt{\pi}} \frac{\Gamma}{\kappa} \sum_{d = 0}^{i}  \frac{\Gamma(d+3/2)}{\Gamma(d + 2)} = \frac{-2}{\sqrt{\pi}} \frac{\Gamma}{\kappa} \left(  \frac{\Gamma(i+5/2)}{\Gamma(i + 2)} - \frac{\sqrt{\pi}}{2} \right) \sim \frac{-2}{\sqrt{\pi}} \frac{\Gamma}{\kappa} \left(  \sqrt{i} - \frac{\sqrt{\pi}}{2} \right).
\end{align}
From here, we can directly calculate the output field as
\begin{align}
\langle \hat a^\dagger_j \hat a_j \rangle &= \frac{2 \Gamma}{ \kappa} - \frac{\langle \hat c_i^\dagger \hat c_i  \rangle}{2j + 4}.
\end{align}
which is asymptotically constant since $\langle \hat c_i^\dagger \hat c_i \rangle$ grows like $\sqrt{j}$.

Finally, we can compute the multi-time steady state correlation functions. Given a round trip time $\tau$, lets assume that $t = n \tau + t'$. In this case, we can write the two-time output field correlation as
\begin{align}
    \langle \hat a^\dagger(t) \hat a(0) \rangle &= \langle [\hat c_{n + 1}^\dagger(t') + \hat c_n^\dagger(t')][\hat c_1(0) + \hat c_0(0)]\rangle
\end{align}

\begin{figure}[t]
    \centering
    \includegraphics{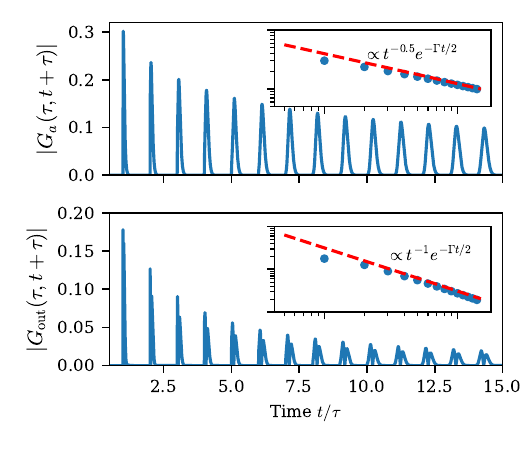}
    \caption{Here, we plot the two-time correlation function (upper) and output field (lower) for the free fermion approximation in the steady state for the same parameters as Fig. 4 in the main text. We can clearly observe the power laws given by $t^{-0.5}$ and $t^{-1}$, respectively, which are distinct from the spin model.}
    \label{fig:free_fermi_output}
\end{figure}

In order to do this, we need to calculate the two-point two-time fermion correlation functions $\langle \hat c_i^\dagger(t') \hat c_j(0) \rangle$ in the steady state. This is given by
\begin{align}
    \langle \hat c_i^\dagger(t') \hat c_j(0) \rangle &= \Gamma \int_0^\infty (G(t' + t) G^\dagger(t))_{i,j} \dd t
\end{align}
Let's begin by calculating $\langle \hat c_n^\dagger(t') \hat c_0(0) \rangle$ as it is the simplest. We have that
\begin{align}
    \langle \hat c_n^\dagger(t') \hat c_0(0) \rangle &= \Gamma \int_0^\infty (G(t' + t) G^\dagger(t))_{n,0} \dd t = \Gamma \int_0^\infty \frac{[-\kappa(t + t')]^n}{n!} e^{-(2\kappa + \Gamma)(t + t'/2)} \dd t \nonumber \\
    &= \frac{\Gamma}{2 \kappa + \Gamma} \left( \frac{-\kappa}{2 \kappa + \Gamma} \right)^n \frac{e^{(\kappa + \Gamma/2)t'}}{n!} \int_{(2 \kappa + \Gamma)t'}^\infty x^n e^{-x} \dd x
\end{align}
where we made the change of variables $x = (2 \kappa + \Gamma)(t + t')$. The remaining integral is an incomplete Gamma function, which we will define as $\Gamma_\mathrm{inc}(y,n) \equiv \int_y^\infty x^n e^{-x} \dd x$. Proceeding in a very similar fashion, we can calculate that
\begin{align}
    \langle \hat c_n^\dagger(t') \hat c_1(0) \rangle &= \Gamma e^{-(\kappa + \Gamma/2)t'} \int_0^\infty  e^{-(2\kappa + \Gamma)t} \left[ -\kappa t' \frac{[-\kappa(t + t')]^n}{n!} +  \frac{[-\kappa(t + t')]^{n-1}}{(n-1)!} \right] \dd t \nonumber \\
    &= \frac{\Gamma}{2 \kappa + \Gamma} \left( \frac{-\kappa}{2 \kappa + \Gamma} \right)^n \frac{e^{(\kappa + \Gamma/2)t'}}{n!} \int_{(2 \kappa + \Gamma)t'}^\infty \left(\kappa t' - \frac{\kappa}{2 \kappa + \Gamma} x \right)x^n e^{-x} \dd x \nonumber \\
    & \ \ \ \  + \frac{\Gamma}{2 \kappa + \Gamma} \left( \frac{-\kappa}{2 \kappa + \Gamma} \right)^{n-1} \frac{e^{(\kappa + \Gamma/2)t'}}{(n-1)!} \int_{(2 \kappa + \Gamma)t'}^\infty x^{n-1} e^{-x} \dd x \nonumber \\
    &= \frac{\Gamma e^{(\kappa + \Gamma/2)t'}}{2 \kappa + \Gamma} \left[ \left( \frac{-\kappa}{2 \kappa + \Gamma} \right)^{n+1} \frac{\Gamma_\mathrm{inc}((2 \kappa + \Gamma)t',n + 1)}{n!}  \right. \nonumber \\
    & \ \ \ \ \left.  +  \kappa t' \left( \frac{-\kappa}{2 \kappa + \Gamma} \right)^{n} \frac{\Gamma_\mathrm{inc}((2 \kappa + \Gamma)t',n)}{n!} + \left( \frac{-\kappa}{2 \kappa + \Gamma} \right)^{n-1} \frac{\Gamma_\mathrm{inc}((2 \kappa + \Gamma)t',n-1)}{(n-1)!} \right]
\end{align}
All together this gives us everything needed to calculate the output field, which is plotted in Fig.~\ref{fig:free_fermi_output}.

\subsection{Many-body Spectrum and Time Dynamics}

Here, we will study the many-body spectrum, and how it relates to the time dynamics. We will begin by considering a critical quench experiment, where the spins are all initialized to the up state $|\uparrow\rangle^{\otimes n}$, and all relax to the down state and there is no pumping $\Gamma = 0$. We can immediately extract the spectral functions from the Green's function. Crucially, the Green's function is non-diagonalizable, so one must be careful when describing its spectrum. The form is that of a single Jordan block, and so the generalized eigenvalues are given by the diagonal entries, $\lambda^{-1} = \omega - \Delta + i \kappa$. We note that this actually gives the entirety of the many-body spectrum, due to the lower triangular nature of the Lindbladian \cite{Torres2014}. This tells us that there is in the entire problem, a single relevant frequency $\Delta$ which has a single relevant relaxation rate $\kappa$.

When the problem in question is Hermitian, we know that the spectrum provides all of the relevant information on the time-dynamics and relaxation. However, the very singular nature of the Green's function actually means that the dynamics can be more complicated, and surprisingly long relaxation time scales can emerge without a corresponding frequency. To see this, let's calculate explicitly the time evolution of the local densities:
\begin{align}
    \langle \hat c_i^\dagger \hat c_i \rangle(t) &= ( G(t) G(t)^\dagger )_{ii} .
\end{align}
where we are using the temporal representation of the Green's function instead of the frequency one. Specifically, we have that
\begin{align}
    G(t) &= \exp\left( i H_{\mathrm{eff}} t  \right) =  \left\{ 
    \begin{array}{cc}
    \exp(- \kappa t) \frac{(-\kappa t)^{i - j}}{(i-j)!} & i \geq j \\
    0 & i < j
    \end{array} \right. ,
\end{align}
which in turn implies that
\begin{align}
    \langle \hat c_i^\dagger \hat c_i \rangle(t) &= e^{-2 \kappa t} \sum_{d = 0}^i \frac{(-\kappa t)^{2d}}{(d!)^2} .
\end{align}
We can directly interpret the exponential prefactor as the light flowing out of a given lattice site, and each term in the sum as all the light flowing in after being emitted by the lattice site $d$. First, let's suppose the lattice site is extremely far from the boundary, and we can extend the sum to infinity. In this case, we can directly calculate that
\begin{align}
    \lim_{i \to \infty} \langle \hat c_i^\dagger \hat c_i \rangle(t) &= e^{-2 \kappa t} \sum_{d = 0}^\infty \frac{(-\kappa t)^{2d}}{(d!)^2} = e^{-2 \kappa t} I_0(2 \kappa t) \sim \frac{1}{\sqrt{4 \pi \kappa t}} .
\end{align}
And so the density decays \textit{algebraically}, without a time scale. This behavior is normally associated with gapless modes, but in this particular problem we know the Green's function is completely gapped. However, the non-Hermiticity of the problem is hiding what is going on. If, instead of looking at the spectrum of the effective Hamiltonian we were to look at its pseudo-spectrum \cite{trefethen2005}, we would indeed find that it has a gap closing in the thermodynamic limit. The problem is that the eigenvectors of the Hamiltonian are all exponentially localized to one side of the chain, which causes a relaxation time that scales with system size, often called the non-hermitian skin effect (NHSE). This anomalous relaxation rate in addition to the algebraic scaling of correlations is direct evidence of this.

Now let's consider a lattice site far down the chain, but not infinitely far away. Here, we can note that (using Stirling's approximation):
\begin{align}
    e^{-2\kappa t} \frac{(-\kappa t)^{2d}}{(d!)^2} &\sim \frac{1}{2 \pi \kappa t} e^{-(\kappa t - d)^2/d} \\
    \implies \langle \hat c_i^\dagger \hat c_i \rangle(t) &\sim \sum_{d = 1}^i \frac{1}{2 \pi \kappa t} e^{-(\kappa t - d)^2/d}
\end{align}
And so each summand is tightly peaked around $\kappa t = i$. Hence, the scaling $1/\sqrt{\kappa t}$ persists up to $\kappa t = d$ after which the population drops exponentially to 0.

It is also interesting to note that algebraic behavior can also be seen in the output field. Note that the output light is captured by
\begin{align}
    \langle \hat a^\dagger_j \hat a_j \rangle &= \langle \hat c_j^\dagger \hat c_j \rangle + \langle \hat c_{j + 1}^\dagger \hat c_j \rangle +  \langle \hat c_{j}^\dagger \hat c_{j + 1} \rangle + \langle \hat c_{j + 1}^\dagger \hat c_{j + 1} \rangle \nonumber \\
    &= 2e^{-2 \kappa t} \sum_{d = 0}^j \frac{(-\kappa t)^{2d}}{(d!)^2} \left(1 - \frac{\kappa t}{d + 1} \right) + e^{-2 \kappa t} \frac{(-\kappa t)^{2j + 3}}{[(j + 1)!]^2}.
\end{align} 
Asymptotically, for large $j$, this scales as
\begin{align}
    \lim_{j \to \infty} \langle \hat a^\dagger_j \hat a_j \rangle &= e^{-2 \kappa t} \left[ I_0(2 \kappa t) - I_1(2 \kappa t) \right] \sim \frac{1}{8 \sqrt{\pi}}(\kappa t)^{-3/2}.
\end{align}

Now, we consider what happens when there is gain in the system. Note that the new effective Hamiltonian is given by
\begin{align}
    H_{\mathrm{eff, pump}} &= H_{\mathrm{eff, no \ pump}} - \frac{i \Gamma}{2} \mathbf{1} 
\end{align}
where $\mathbf{1}$ is the identity matrix. This implies that the spectrum is uniformly shifted down in the complex plane by $\Gamma/2$, and so what was originally a gapless pseudo-spectrum is now gapped. This would lead to the intuition that the algebraic behavior will be cutoff on a time-scale $\Gamma^{-1}$, which we will see subsequently.

The Green's function are modified to take the form (in the time domain):
\begin{align}
    G_{ij}(t) &= \exp(i H_{\mathrm{eff}} t) = \left\{
    \begin{array}{cc}
    e^{-(\kappa + \Gamma/2)t} \frac{(- \kappa t)^{i - j}}{(i-j)!}     & i \geq j \\
      0   & i < j
    \end{array}
    \right. .
\end{align}
Defining the covariance matrix $C = \langle \hat c_i^\dagger \hat c_j \rangle$, the dynamics are given by
\begin{align}
    C(t) &= \Gamma \int_0^t G(\tau) G^\dagger(\tau) \dd \tau + G(t) C(0) G^\dagger(t) .
\end{align}
Taking an initial condition of all spins pointing up implies that $C(0) = \mathbf{1}$, allowing the simplification:
\begin{align}
    C(t) &= \Gamma \int_0^t G(\tau) G^\dagger(\tau) \dd \tau + G(t) G^\dagger(t) .
\end{align}
Now, we can calculate that 
\begin{align}
    (G(t) G^\dagger(t))_{ii} &= e^{-(2 \kappa + \Gamma) t}  \sum_{d = 0}^i \frac{(-\kappa t)^{2d}}{(d!)^2} .
\end{align}
which scales asymptotically as
\begin{align}
    \lim_{i \to \infty, t \to \infty}(G(t) G^\dagger(t))_{ii} &= \frac{ e^{-\Gamma t} }{\sqrt{4 \pi \kappa t}} .
\end{align}
As expected, up to times of order $\Gamma^{-1}$ the scaling is algebraic, before being cutoff to saturate at a steady state value due to the pumping. The wavefront is still visible for spins that don't reach the steady state value before this time scale. I.e., this gives us an effective length scale after which the shock wave is lost given by $\kappa/\Gamma$, which is unsurprisingly the same length scale that appears in the steady state distribution.

Similarly, we can calculate that the output field obeys
\begin{align}
    \lim_{i,t \to \infty} \langle \hat a^\dagger_i \hat a_i \rangle &=  \frac{e^{-\Gamma t}}{8 \sqrt{\pi}}(\kappa t)^{-3/2} + \frac{\Gamma}{2 \kappa} \left( 1 - \frac{e^{-\Gamma t}}{\sqrt{4 \pi \kappa t}}\right) + \mathcal{O}(\Gamma/\kappa)^2
\end{align}
This is a different power law than the total density, but it still decaying algebraically. The change in power law results from the fact that there is some destructive interference between the output field at time $j\tau$ and $(j+1)\tau$.

\subsection{Additional Dissipation}

Two extremely relevant sources of imperfections in a possible experiment both would to photon loss: either losing photons from the waveguide during transmission, or through qubit $T_1$ relaxation. As it turns out, these end up being (up to some trivial rescaling) completely equivalent. To see this, let's begin by first considering simple $T_1$ relaxation, which we model via the Lindblad jump operator:
\begin{align}
    \hat L_{i,\mathrm{inc}} &= \sqrt{\gamma_L} \hat \sigma_i^- .
\end{align}
We can calculate exactly the affect this has on the single particle Green's function, which is modified to be:
\begin{align} 
G_{ij}(\omega) &= (\omega - H_{\mathrm{eff}} ) ^{-1}_{ij} = \frac{1}{\chi} \times \left\{ 
\begin{array}{cc}
\left(- \frac{i \kappa}{\chi} \right)^{i - j} & i \geq j \\
0 & i < j
\end{array} \right. , \\
\chi &= \omega - \Delta + i\left(\kappa + \frac{\Gamma + \gamma_L}{2}\right),
\end{align}
effectively broadening the resonance, as one might expect. This also implies that the cutoff length scale when we are no longer able to see the algebraic scaling will be
\begin{align}
    \xi_L^{-1} &= -\log \left( \frac{2 \kappa}{2 \kappa + \Gamma + \gamma_L} \right) \sim \frac{\Gamma + \gamma_L}{2 \kappa} .
\end{align}
On the other hand, we can model waveguide loss using the SLH formalism \cite{Combes2017}. Suppose the probability a photon survives an entire round trip of the waveguide is given by $\eta^2$. Then we can rewrite the master equation as
\begin{align}
\hat H &= \frac{\Delta}{2} \sum_{i = 1}^N \hat \sigma_i^z - \frac{i \eta \kappa}{2} \sum_{i = 1}^{N-1} \hat \sigma_i^x \hat \sigma_{i + 1}^x + \hat \sigma_i^y \hat \sigma_{i + 1}^y , \\
\partial_t \hat \rho &= -i[\hat H, \hat \rho] + \Gamma \sum_{i = 1}^N \D[\hat \sigma^+_i] \hat \rho + \kappa \sum_{i = 0}^N \D[\eta \hat \sigma_i^- + \hat \sigma^-_{i + 1}] \hat \rho + \kappa (1 - \eta^2) \sum_{i = 0}^N \D[ \hat \sigma_i^-] \hat \rho .
\end{align}
We can once again recalculate the single-particle Green's function analytically to find that 
\begin{align} 
G_{ij}(\omega) &= (\omega - H_{\mathrm{eff}} ) ^{-1}_{ij} = \frac{1}{\chi} \times \left\{ 
\begin{array}{cc}
\left(- \frac{i \kappa'}{\chi} \right)^{i - j} & i \geq j \\
0 & i < j
\end{array} \right. , \\
\chi &= \omega - \Delta + i\left(\kappa' + \frac{\Gamma}{2} + \frac{1 - \eta}{\eta} \kappa' \right) , \\
\kappa' &= \eta \kappa ,
\end{align}
where we defined the rescaled coupling constant $\kappa'$ that takes into account the fact that not all of the light will survive the waveguide. Note that, up to a rescaling of $\kappa$, this is of the exact same form as before, where we identify $\gamma_L = \frac{2(1-\eta)}{\eta} \kappa$, which gives a rescaled length:
\begin{align}
    \xi_L^{-1} &= \frac{\eta}{1 - \eta} + \frac{\Gamma}{2 \eta \kappa}.
\end{align}
Thus, we can see that these two effects appear identically in the experiment. Besides changing the cutoff length scale $\xi_L$, it will also overall suppress the steady state density by a constant factor, but this will not change the actual functional form, only rescale it by a constant prefactor. 

\section{Reset}

\begin{figure}
\centering 
\includegraphics[width=4.5in]{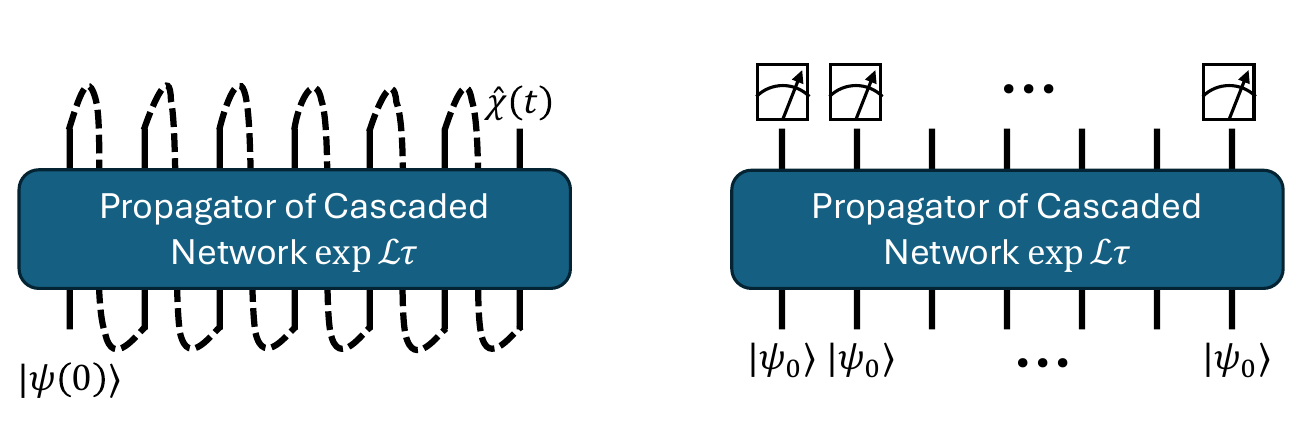}
\caption{Tensor-network depiction of the cascaded network. Dashed lines show tensor contractions between $|\psi_i(\tau) \rangle$ and $\psi_{i+1}(0)\rangle$ in the left. This contractions are broken by measurement in the right and fresh spins are fed in.}
\label{fig:reset}
\end{figure}

Here we consider qubit reset. Specifically, we are interested in practical experimental considerations of the fast reset protocol. Recall from the main text that in order to perform a quench experiment to study short time dynamics, it is important to be able to prepare an initial unentangled state. This is summarized in Fig.~\ref{fig:reset}, where we can compare the case where we either don't reset (left) or do reset (right) the qubit at each round trip time $\tau$. On the left, when we do nothing, the qubit tensor $|\psi_i(\tau)\rangle$ is contracted with $|\psi_{i + 1}(0)\rangle$ (dashed line), signifying that they are in fact the exact same thing. In order for them to be different, we perform a projective measurement at time $\tau$ and feed in a fresh state $|\psi_0\rangle$.

As written, this process occurs instantaneously. In practice, it must be significantly faster than the time scale of the interaction with the waveguide mode. I.e., the reset time $t_{\mathrm{reset}} \ll \kappa^{-1}$. If we are strongly coupled to the waveguide, a superconducting qubit can often be limited by it's measurement time, and so this could prove difficult experimentally. In order to fix this, we consider the following scenario. Firstly, let's assume that the qubit frequency exists far outside the pass band of the metamaterial waveguide, and so they are generically not interacting at all to first order. We will instead assume that the coupling is activated by a sideband drive, and so we can control when the atom/waveguide coupling is on and off. In this case, we can decouple the waveguide from the qubit at time $t = \tau - t_\mathrm{reset}$. Then we measure and reset the qubit and recouple it at exactly $\tau$. This means the maximal simulation time allowed goes from $\tau \to \tau - t_\mathrm{reset}$, so in principle it will also be important that $\tau \gg t_\mathrm{reset}$ to get a long simulation period.

\section{Measuring observables from the output field}

In Table I of the main text, we give an expression of the onsite density $\langle \sigma^+_n \sigma^-_n\rangle_t$ in terms of correlation functions of the output field $b_{\rm out}(t)$ of the waveguide (i.e., the field emitted to the left in Fig.~1(a) of the main text).
Here, we generalize this to all two-point correlation functions of the many-body chain.
Via the standard input-output relation, the output field of the single QE system is
$b_\mathrm{out}(t) = \sqrt{\kappa}[\sigma^{-}(t) + e^{-\mathbbm{i}\phi}\sigma^{-}(t-\tau)] + e^{-\mathbbm{i}\phi}b_{\rm in}(t)$, where $\phi=\pi-\omega_0\tau$ is the phase acquired by a photon during a round-trip in the delay loop and $b_{\rm in}(t)$ is the input field, which obeys the field commutator $[b_{\rm in}(t),b_{\rm in}^\dagger(t')] = \delta(t-t')$.
The system operators $\sigma^-(t)$ map to the many-body lattice operators via $\sigma_n^-(t) = \sigma^-[(n-1)\tau + t]$, for $t\in[0,\tau]$, and we find it useful to map to the many-body output fields accordingly, $b_{{\rm out},n}(t) = b_{\rm out}[(n-1)\tau + t]$.
Thus, $b_{{\rm out},n}(t) = \sqrt{\kappa}[\sigma^{-}_n(t) + e^{-\mathbbm{i}\phi}\sigma^{-}_{n-1}(t)] + e^{-\mathbbm{i}\phi}b_{{\rm in},n}(t)$, where now $[b_{{\rm in},n}(t),b_{{\rm in},n'}^\dagger(t')] = \delta_{nn'}\delta(t-t')$.
We can therefore express $\sigma_n^-(t)$ in terms of the input and output fields of all lattice sites $j\leq n$:
\begin{align}
    \sqrt{\kappa}\sigma^-_n(t) = \sum_{j=0}^{n-1} (-1)^j e^{-\mathbbm{i}j\phi}[b_{{\rm out},n-j}(t) - e^{-\mathbbm{i}\phi}b_{{\rm in},n-j}(t)].
\end{align}
From this expression, we immediately see that correlation functions of the many-body chain can be obtained from specific linear combinations output field correlations. E.g., assuming the waveguide is initially in vacuum $\langle b_{{\rm in},n}^\dagger(t) b_{{\rm in},n'}(t')\rangle = 0$, the standard beamsplitter correlations are
\begin{align}
    \kappa \langle \sigma^+_n(t)\sigma^-_m(t')\rangle = \sum_{j=0}^{n-1}\sum_{k=0}^{m-1} (-1)^{j+k}e^{-\mathbbm{i}(j-k)\phi}\langle b_{{\rm out},n-j}^\dagger(t) b_{{\rm out},m-k}(t')\rangle.
\end{align}
This is precisely the expression given in Table I of the main text for $t'=t$.
Here, the output field correlations are the unnormalized first order coherence functions $G^{(1)}[(n-j-1)\tau+t,(m-k-1)\tau+t']$.
Similarly, anomalous correlations of the many-body system $\langle \sigma^-_n(t)\sigma^-_m(t')\rangle$ are found by measuring the appropriate anomalous correlations of the output field $\langle b_{{\rm out},n-j}(t)b_{{\rm out},m-k}(t')\rangle$ in an analogous expression.
Finally, density-density correlations $\langle\sigma^z_n(t)\sigma^z_m(t')\rangle$ and other four-point functions are computed from the appropriate linear combination of (unnormalized) second order coherence functions $G^{(2)}$ of the output field.

\section{Analysis of a superconducting metamaterial waveguide realization}

Here we discuss the design requirements of a superconducting metamaterial waveguide whose slow propagation velocity (large group index $n_g = c/v_g \gg 1$) allows for long roundtrip times compared to the QE-waveguide coupling, $\tau \gg 1/\kappa$.
Here, we assume the QE to be a transmon qubit and $\omega_0$ to be the transition frequency between its lowest two levels $|g\rangle$ and $|e\rangle$.
Because metamaterial waveguides have nonlinear dispersion, $\omega(k) \neq v_p k$, the roundtrip time is a function of frequency $\tau = \tau(\omega)$, so the integrity of a wavepacket with finite bandwidth $\Delta\omega\sim\kappa$ degrades as it propagates along the waveguide.
Therefore, we must balance the need for a high group index $n_g \gg 1$ at $\omega_0$ with the need for a sufficiently linear dispersion within a bandwidth $\Delta\omega\sim\kappa$ around $\omega_0$.

To estimate the effects of nonlinear dispersion, we need the roundtrip propagation time through a waveguide of length $L$ as a function of frequency
\begin{align}
    \tau(\omega) = \frac{2L}{v_g(\omega)},
\end{align}
where $v_g = \frac{\partial}{\partial k}\omega(k)$ is the group velocity.
We expand the dispersion $\omega(k)$ about the center wavevector $k_0$ defined by $\omega(k_0) = \omega_0$:
\begin{align}
    \omega(k-k_0) = \omega_0 + v_0 (k-k_0) + \frac{1}{2}\alpha(k-k_0)^2 + \frac{1}{6}\beta(k-k_0)^3,
\end{align}
where $v_0$ is the group velocity at $\omega_0$, and the two lowest order nonlinearities are $\alpha$ and $\beta$.
From this, we derive the group velocity as a function of wavevector $v_g(k-k_0) = \frac{\partial\omega}{\partial k} = v_0 +\alpha(k-k_0)+\frac{1}{2}\beta(k-k_0)^2$. 
Inverting the approximate dispersion to $\mathcal{O}(\omega-\omega_0)^3$, we find the group velocity as a function of frequency
\begin{align}
    v_g(\omega-\omega_0) = v_0 \left[ 1 + \frac{\alpha}{v_0^2}(\omega-\omega_0) + \frac{1}{2}\left( \frac{\beta}{v_0^3} - \frac{\alpha^2}{v_0^4} \right) (\omega - \omega_0)^2 \right] + \mathcal{O}(\omega - \omega_0)^3,
\end{align}
then, we find the roundtrip time to second order in $(\omega-\omega_0)$ to be
\begin{align}
    \tau(\omega-\omega_0) = \frac{2L}{v_0}\left[ 1 - \frac{\alpha}{v_0^2}(\omega-\omega_0) + \left( \frac{\alpha^2}{v_0^4} -  \frac{\beta}{2v_0^3} \right) (\omega - \omega_0)^2 \right].
\end{align}
With this expression in hand, we estimate the RMS spread in the roundtrip time of a wavepacket with bandwidth $\Delta\omega\sim\kappa$: $\Delta\tau_{\rm rms} = \sqrt{\langle\tau^2\rangle - \langle\tau\rangle^2}$, where the average $\langle\cdot\rangle$ is taken over $(\omega-\omega_0)\in [-\kappa,\kappa]$.
We find
\begin{align}
    \Delta\tau_{\rm rms} = \tau_{0}\sqrt{\frac{1}{3}\frac{\alpha^{2}}{v_{0}^{4}}\kappa^2+\frac{4}{45}\left(\frac{\alpha^{2}}{v_{0}^{4}}-\frac{\beta}{2v_{0}^{3}}\right)^{2}\kappa^{4}},
\end{align}
where $\tau_0 = 2L/v_0$ is the mean roundtrip time.
The effects of nonlinear dispersion are negligible when $\Delta\tau_{\rm rms}$ is much smaller than the interaction time with the QE: $\kappa\Delta\tau_{\rm rms} \ll 1$.

We take the metamaterial waveguide design of Ref.~\cite{mirhosseini_Superconducting_2018} as our prototype.
The design is a coplanar waveguide periodically loaded with lumped-element LC resonators.
The details of the lumped-element model and physical design can be found in the Supplementary Information for Ref.~\cite{mirhosseini_Superconducting_2018}.
Using the second waveguide design fabricated in that work, we find that at center frequency $\omega_0 = 2\pi\times 4.8~\text{GHz}$ (near the band edge of the lower transmission band), the waveguide has a group index $n_g \approx 1100$.
(This is computed from the analytic expression of the waveguide dispersion $\omega(k)$ derived from the lumped-element model, see the Supplementary Information for Ref.~\cite{mirhosseini_Superconducting_2018}.)
If we couple a $4.8~\text{GHz}$ transmon qubit to the waveguide with rate $\kappa = 2\pi\times 2~\text{MHz}$, and design a $12~\text{cm}$ section of waveguide ($\approx 342$ unit cells) between the transmon and a reflective termination, the mean roundtrip time is $\kappa\tau_0 \approx 11.5$ and the RMS dispersion is $\kappa \Delta\tau_{\rm rms}\approx 0.10$.
This implies that the steady state properties of a length $N\lesssim 11$ chain can be probed with this system, so long as the loss in the waveguide is sufficiently small.

As discussed in the main text, loss in the waveguide contributes to the effective intrinsic loss of the qubit.
Following Ref.~\cite{mirhosseini_Superconducting_2018}, we model the waveguide losses as resistances $R_0$ and $R_r$ in series with the respective inductors.
The effective shunt impedance of each unit cell is the parallel sum (i.e., $a||b = [a^{-1}+b^{-1}]^{-1}$) of the shunt impedance of $C_0$ and the impedance of the capacitively-coupled resonator at $\omega_0$:
\begin{align}
    Z_{\rm eff} &= \frac{1}{i\omega_0 C_0} \Big|\Big| Z_r
    \\
    Z_r &= \frac{1}{i\omega_0 C_g} +  \big(R_r + i\omega_0 L_r\big) \Big|\Big| \frac{1}{i\omega_0C_r}.
\end{align}
We find that the effective reactance is capacitive, $X_{\rm eff} = {\rm Im} Z_{\rm eff} = -14.3~\Omega$; therefore, we can define an effective shunt capacitance $C_{\rm eff}$ and shunt inductance $G_{\rm eff}$ (valid only in a small bandwidth $\Delta\omega\sim\kappa$ around $\omega_0$) to reduce the full lumped-element model to an effective simple transmission line model with unit cells consisting of a series inductance $L_0$ and resistance $R_0$ and a shunt capacitance $C_{\rm eff}$ and conductance $G_{\rm eff}$ (see e.g. \cite{pozar_microwave_2012}).
Following the standard analysis of low loss transmission lines (i.e., for which $R\ll \omega_0 L$ and $G \ll \omega_0 C_{\rm eff}$), we find the attenuation coefficient
\begin{align}
    \alpha_{\rm wg} = \frac{1}{2}\left( \frac{R_0}{Z_0} + G_{\rm eff}Z_0 \right),
\end{align}
where $Z_0 = \sqrt{L_0/C_{\rm eff}}$ is the effective characteristic impedance \cite{pozar_microwave_2012}.
We find $Z_0 = 1.58~\Omega$ and $\alpha = 2.03\times10^{-4}$; the latter is the attenuation per unit cell.
The round-trip attenuation factor of the $12~\text{cm}$ waveguide is $\exp(-\alpha_{\rm wg}N_{\rm cell})\approx 0.870$, for $N_{\rm cell} = 2\times342$ unit cells.
The effective loss rate is found by computing the $1/e$ attenuation distance $\delta l_{\rm att} = \alpha_{\rm wg}^{-1}\times 350$~\textmu m.
Thus, the effective loss rate is $\gamma_{\rm L} = v_g/\delta l_{\rm att} \approx 2\pi\times 24.4$ kHz.
This is equivalent to a qubit lifetime of $T_1 = 6.5$ \textmu s.
Assuming this is the dominant source of parasitic dissipation, it limits the simulable chain length to $N_{\rm max}\sim \sqrt{\kappa/\gamma_{\rm L}} \approx 9$.
Therefore, we conclude that this metamaterial waveguide could be used to probe the steady state properties of a nonreciprocal chain with a maximum length $N\sim 10$.

\bibliography{references}